\documentstyle [12pt]{article}
\newcommand{\pbarp}{\overline{p}p}
\newcommand{\qqbar}{Q\overline{Q}}
\newcommand{\KKbar}{K\overline{K}}
\newcommand{\pbarn}{\overline{p}n}
\newcommand{\nnbar}{\overline{n}n}
\newcommand{\uubar}{u\overline{u}}
\newcommand{\ddbar}{d\overline{d}}
\newcommand{\ssbar}{s\overline{s}}

\newcommand{\bea}{\begin{eqnarray}}
\newcommand{\eea}{\end{eqnarray}}
\begin{document}
\title{\small \rm \begin{flushright} \small{hep-ph/9507326}\\
\small{RAL-TR-95-003}\\
14 July 1995 \end{flushright} \vspace{2cm}
\LARGE {\bf Is $f_0(1500)$ a Scalar Glueball?}
 \vspace{0.8cm} }
\author{Claude Amsler\thanks{E-mail : amsler@cernvm.cern.ch}\\
{\small \em Physik-Institut, Universit\"at Z\"urich, CH-8057
Z\"urich,
Switzerland} \\ \\
 Frank E. Close\thanks{E-mail : fec@v2.rl.ac.uk} \\
{\small \em Particle Theory, Rutherford-Appleton Laboratory,
Chilton,
Didcot OX11 0QX, UK} \\  \\}
\date{July 1995 \vspace{1.0cm}}

\begin{center}
\maketitle

\begin{abstract}
Following the discovery of two new scalar mesons, $f_0(1370)$
and $f_0(1500)$ at the Low Energy Antiproton Ring at CERN,
we argue that the observed properties of this pair are
incompatible with them both being  $\qqbar$
mesons. We show instead that $f_0(1500)$ is compatible with
the ground state glueball expected around 1500 MeV mixed with
the nearby states of the $0^{++}$ $Q\bar{Q}$ nonet. Tests of this
hypothesis include the prediction of a further scalar state,
$f_0'(1500 - 1800)$ which couples strongly to $K\bar{K}$,
$\eta\eta$ and $\eta\eta'$. Signatures for a possible tensor glueball
at $\sim $2 GeV are also considered.
\end{abstract}
Submitted to Phys. Rev. D
\end{center}

\newpage

\section{Introduction}

Glueballs are a missing link of the standard model.
Whereas the gluon degrees of freedom expressed
in $L_{QCD}$ have been established beyond doubt in high
momentum data, their dynamics in the strongly interacting limit
epitomised by hadron spectroscopy are quite obscure. This may be
about to change as a family of candidates for gluonic hadrons (glueballs and
hybrids) is now emerging \cite{amsler94,cp94}. In this paper we
shall argue that scalar mesons around 1.5 GeV, in particular the
detailed phenomenology of $f_0(1500)$ and its partner
$f_0(1370)$, suggest that a glueball exists in
this region, probably mixed with nearby isoscalar members of the
scalar nonet. This hypothesis may be tested in forthcoming
experiments.

In advance of the most recent data, theoretical arguments
suggested that there may be gluonic activity manifested in the 1.5
GeV mass region. Lattice QCD is the best simulation of theory and
predicts the lightest ``primitive" (ie quenched approximation)
glueball to be $0^{++}$ with mass $1.55 \pm 0.05$ GeV
\cite{ukqcd}. Recent lattice computations place the glueball
slightly higher in mass at $1.74 \pm 0.07$ GeV \cite{weing}
with an optimised value for phenomenology proposed by Teper\cite{teper}
of $1.57 \pm 0.09$ GeV. That lattice QCD computations of the scalar glueball
mass are now concerned with such fine details represents considerable
advance in this field. Whatever the final concensus may be, these
results suggest that scalar mesons in
the 1.5 GeV region merit special attention.
Complementing this has been the growing realisation that there
are now too many $0^{++}$ mesons confirmed for them all to be
$\qqbar$ states \cite{amsler94,close92,PDG}.

At $\sim 1.5$ GeV there is a clear $0^{++}$ signal, $f_0(1500)$, in
several experiments  \cite{Anis}-\cite{Gentral}, whose serious
consideration for being associated with the primitive
glueball is enhanced by the fact that its
production is by mechanisms traditionally believed to be those
that favour gluonic excitations. Specifically these include
\cite{closerev}
\begin{enumerate}
\item
Radiative $J/\psi$ decay: $J/\psi  \rightarrow \gamma+G$
\cite{bugg}
\item
Collisions in the central region away from quark beams and target:
$pp \rightarrow p_f(G)p_s$ \cite{Kirk,Gentral}.
\item
Proton-antiproton annihilation where the destruction of quarks
creates opportunity for gluons to be manifested. This is the Crystal
Barrel \cite{Anis}-\cite{Enhan} and E760
\cite{Hasan1,Hasan2} production mechanism in which detailed
decay systematics of $f_0(1500)$ have been studied.
\item
Tantalising further hints come from the claimed sighting \cite{had95}
of the $f_0(1500)$ in decays of the hybrid meson candidate \cite{cp94}
$\pi(1800) \rightarrow \pi f_0(1500) \rightarrow \pi \eta \eta$.
\end{enumerate}

The signals appear to be prominent in decay channels such as
$\eta\eta$ and $\eta\eta'$ that are traditionally regarded as
glueball signatures. However, such experiments are
not totally novel and some time ago one of us (FEC)
addressed the question of why glueballs had remained hidden
during 25 years of establishing the Particle Data Group list
\cite{PDG} of $\qqbar$ states. This was suggested \cite{closerev}
to be due to the experimental concentration on a restricted class of
production mechanisms and on final states
with charged pions and kaons. The more recent emphasis on
neutral final states (involving $\pi^0$, $\eta$, $\eta'$) was
inspired by the possibiblity that $\eta$ and $\eta'$ are strongly
coupled to glue. This dedicated study of neutrals was a new
direction pioneered by the GAMS Collaboration at CERN
announcing new states decaying to $\eta\eta$ and
$\eta\eta'$ \cite{Aldeall}.

The Crystal Barrel collaboration at LEAR has made intensive study
of $p\bar{p}$ annihilation into neutral final states involving
$\pi^0$, $\eta$ and $\eta'$. They find a clear signal
for $f_0(1500)$ in $\pi^0\pi^0$, $\eta \eta$ and $\eta \eta'$.
Our present work extends and generalises the work of ref. \cite{closerev}
in the light of these new data from LEAR. Our purpose
is to examine the data on the $f_0(1500)$, compare with
predictions for glueballs and identify the seminal experiments
now needed to confirm that gluonic degrees of freedom are being
manifested in this region. A summary of this work has already
been published elsewhere \cite{letcafe}.

The structure of the paper is as follows. We first review the
experimental data on scalar mesons with special emphasis on
states seen in the Crystal Barrel detector at LEAR. We then derive
from SU(3)$_f$ the branching ratios for $\qqbar$ decays into two
pseudoscalars, show that this successfully describes the known
decay rates in the $2^{++}$ $\qqbar$ nonet, and then compare our
predictions to the observed decay modes of $f_0(1500)$.
Previous bubble chamber experiments have not observed a
$\KKbar$ signal in the 1500 MeV mass region which, if confirmed, would imply
a set of branching ratios that are unnatural for a state belonging
to a quarkonium nonet. If a significant signal were to be observed, it would
be possible to find a quarkonium mixing angle that reproduces the
observed final state abundances; however
the systematics would then imply that $f_0(1500)$ is
dominantly $n\bar{n} \equiv (u\bar{u} + d\bar{d})/\sqrt{2}$. This
would have two immediate consequences:
\begin{enumerate}
\item
This would leave the $f_0(1370)$ state, which is also seen in
$p\bar{p}$ annihilation with decay branching ratios and total width consistent
with an $n\bar{n}$ structure
\cite{Amsler3pi0,Amsleretaeta}, isolated.
\item
With either the $f_0(1370)$ or $f_0(1500)$ assigned as the $n\bar{n}$ member,
the orthogonal quarkonium in the nonet would have to be
dominantly $s\bar{s}$, hence probably heavier than $f_0(1500)$ and
decaying strongly into $K\bar{K}$.   Identification of this state is
now imperative in order to complete the multiplet and discriminate among
hypotheses.
\end{enumerate}
We then show that that the decay rates of $f_0(1500)$ are
compatible with a glueball state whose mass lies
between the $n\bar{n}$ and $s\bar{s}$ scalar
quarkonium states, and whose nearby presence disturbs the
glueball decays in a characteristic flavour dependent manner. In
the climax of the paper we show that dynamics inspired by lattice
QCD may be consistent with the data and we consider the
implications for glueballs mixing with quarkonia in the 1500 MeV
range. We finally show that a reasonable nonet can be constructed
with the remaining scalar mesons.

\section{Scalar Mesons in the Crystal Barrel}
The lowest lying $0^{++}$ mesons, namely the isospin $I$ = 0
$f_0(980)$ and the $I$ = 1 $a_0(980)$, have been assumed to
be $\KKbar$ molecules \cite{wein,cik}. This belief is motivated by
their strong couplings to $\KKbar$ - in spite of their
masses being at the $\KKbar$ threshold - and their small
$\gamma\gamma$ partial widths. For $f_0(980)$,
$\Gamma_{\gamma\gamma}$ = 0.56 $\pm$ 0.11 keV \cite{PDG}.
For $a_0(980)$, one finds with a LEAR measurement of the
relative branching fraction for a$_0$ decay to $\KKbar$ and
$\eta\pi$ \cite{Spanier} the partial width
$\Gamma_{\gamma\gamma}$ = 0.33  $\pm$ 0.13 keV
\cite{amsler94}. Thus the $\gamma\gamma$ partial widths
appear to be nearly equal, close to predictions for $\KKbar$
molecules (0.6 keV) and much smaller than for $\qqbar$ states
\cite{Barnes}.

The nature of these states is likely to be illuminated soon at
DA$\phi$NE \cite{cik}. If they are not simply $\qqbar$ then the $0^{++}$
$\qqbar$ mesons need to be identified. A new  $J^{PC}(I^G) = 0^{++}
(1^-)$ meson, $a_0(1450)\rightarrow\eta\pi$,  has been reported
by the Crystal Barrel collaboration at LEAR \cite{Spanier}. This
state, with a mass of 1450 $\pm$ 40 MeV and a width of 270
$\pm$ 40 MeV, appears, together with $a_0(980)$, in the
$\eta\pi$ S-wave in $\pbarp$ annihilation at rest into
$\eta\pi^0\pi^0$. We shall show in section 8 that
$a_0(1450)$ can be identified with the $I$=1 member of the
ground state $0^{++}$ $Q\bar{Q}$ nonet. In turn this and the
$K^*(1430)$ set the natural energy scale for the scalar nonet.

Recent data in $\pbarp$  annihilation at LEAR into
$\eta\pi^0\pi^0$ \cite{Spanier}, 3$\pi^0$ \cite{Anis,Amsler3pi0}
and $\eta\eta\pi^0$ \cite{Augustin,Amsleretaeta}
require an $I=0$ scalar resonance in the range 1320 to 1400 MeV,
decaying to $\pi^0\pi^0$ and $\eta\eta$. We shall use 1360 MeV
as average mass but shall adopt the nomenclature of the Particle
Data Group \cite{PDG} calling this state $f_0(1370)$. Its width
varies between 200 and 700 MeV, depending on theoretical
assumptions. For example, the 3$\pi^0$ data give $\Gamma\sim$
700 MeV \cite{Amsler3pi0}, decreasing to 300 $\pm$ 80 MeV if
the 700 MeV broad background structure \cite{Au,Morgan}
centered at 1000 MeV in the $\pi\pi$ S-wave (called $f_0(1300)$
in the latest issue of the Particle Data Group) is introduced in the
analysis. This is in good agreement with a coupled channel
analysis of $\eta\pi^0\pi^0$, 3$\pi^0$ and $\eta\eta\pi^0$ which
leads to a mass of 1390 $\pm$ 30 MeV and a width of 380 $\pm$
80 MeV \cite{coupled}. The $\gamma\gamma$ width in this
region is also consistent with it containing the $^3P_0(Q\bar{Q})$
state \cite{MP,BCL}.

There is rather general agreement that the ground state $n\bar{n}$ state
is manifested here. The debate is one of detail on the relationship of
the $f_0(1370)$ to the broad $f_0(1300)$, in particular as to whether these
are two independent states or manifestations of a single state,
 and to what extent unitarity corrections are
important \cite{nt}. This issue is peripheral to our main analysis which
will rely only on the generally accepted association of the ground state
$n\bar{n}$ as the seed for the phenomenology in the $f_0(1300-1370)$
region and that this is distinct from the $f_0(1500)$ state observed by
the Crystal Barrel collaboration decaying to $\pi^0\pi^0$
\cite{Anis,Amsler3pi0}, $\eta\eta$
\cite{Augustin,Amsleretaeta} and $\eta\eta'$ \cite{Enhan}.
This state was seen in $\pbarp$ annihilation into $3\pi^0$,
$\eta\eta\pi^0$ and $\eta\eta'\pi^0$, leading to six final state
photons. The masses and widths observed in the three decay
channels are consistent, giving the average:
\begin{equation}
(m, \Gamma) = (1509 \pm 10, 116 \pm 17) \ {\rm MeV},
\end{equation}
while the coupled channel analysis \cite{coupled} gives 1500
$\pm$ 10 MeV and the less precise but compatible width of 154
$\pm$ 30 MeV. It is possible that $f_0(1500)$ has also been seen
by MARKIII and DM2 in $J/\psi \rightarrow \gamma + \pi \pi \pi
\pi$, hitherto misidentified as $0^{-+}$ \cite{bugg} and in
$J/\psi \rightarrow \gamma \pi\pi$\cite{chen91}. A resonance
decaying to $\pi^0\pi^0$  and $\eta\eta$ was also reported by
E760 in $\pbarp$ annihilation at higher energies at the Fermilab
accumulator, with masses and widths (1508, 103) MeV
\cite{Hasan1} and (1488, 148) MeV \cite{Hasan2}, respectively. A
spin-parity analysis is in progress \cite{Slovenia}.

  The GAMS Collaboration at CERN
\cite{Aldeall} reports a $0^{++}$ 180 MeV broad resonance,
$f_0(1590)$, decaying to $\eta\eta'$, $\eta\eta$ and $4\pi^0$,
also observed in central production \cite{Gentral} and by VES at
Serpukhov \cite{Bela} in $\eta\eta'$: this
might be the $f_0(1500)$ state though the status of the $\pi \pi$
branching ratio needs to be clarified.\footnote{The
$\pi\pi$ branching ratio is the largest deviation but is
``not in contradiction" with the $f_0(1590)$ of GAMS and the
$f_0(1500)$ of Crystal Barrel being the same state\cite{proko95}} A
strong coupling of $f_0(1500)$ to pions would contradict it being
primarily an $s\bar{s}$ state and, as we shall argue later, it is a
candidate for a glueball mixed with the $Q\bar{Q}$ nonet, where
$f_0(1370)$ is dominantly $n\bar{n}$, and a more massive
$s\bar{s}$ remains to be identified.
The $f_0(1500)$, clearly established in different decay
channels and with detailed information on branching ratios to
several channels, will form the fulcrum of our investigation.

There are candidates for this $s\bar{s}$ state though their
existence and/or $s\bar{s}$ assignment remain to be established.
A $0^{++}$ structure, $f'_0(1525)$, with poorly known width
($\sim$ 90 MeV) is observed to decay into $K_SK_S$  in $K^-
p$ interactions \cite{Lass}. This state requires confirmation from
other experiments.
The $\theta(1690)$ (now known as
$f_J(1720)$ \cite{PDG}) is a candidate due to
its affinity for $K\bar{K}$ and $\eta\eta$ decays though its quantum
numbers, $0^{++}$ or $2^{++}$, are still controversial.

An $I=0$ scalar with a width of 56 MeV, is observed  at
1446 MeV by the WA91 collaboration at CERN in pp central
collisions \cite{Kirk}. It decays to four pions, dominantly through
$\rho^0(\pi^+\pi^-)_P$ where the dipion is in a P-wave. This may
be the same as $f_0(1500)$ produced in the second of the
favoured glueball mechanisms (section 1), with its apparent small
width being the result of interference between $f_0(1370)$ and
$f_0(1500)$ \cite{WA91prep}. In any event it does not detract
from the qualitative observation that there are too many
isoscalars observed in various production mechanisms for them all
to be explained naturally within a $Q\bar{Q}$ picture. The fact
that there does not appear to be such copious activity in the $I =
1$ and strange sectors adds weight to the suspicion that glueball
excitation is affecting the $I$ = 0 spectrum.

A substantial part of this paper will examine what the flavour
content of two-body decays can reveal about the structure of the
initial meson. Based on this analysis we shall argue that the
Crystal Barrel $f_0(1500)$ has decay properties incompatible with
a $\qqbar$ state. In addition we shall show that a reasonable
$\qqbar$ nonet may be constructed with $a_0(1450)$,
$f_0(1370)$, $K_0^*(1430)$, and an $\ssbar$ state above 1500
MeV, thus leaving $f_0(980)$ and $f_0(1500)$ as exotic (not simply
$\qqbar$) states.

\section{Quarkonium Decay Amplitudes}
Consider a quarkonium state
\begin{equation}
|\qqbar\rangle = {\rm cos}\alpha |n\bar{n}\rangle -
{\rm sin}\alpha |s\bar{s}\rangle
\end{equation}
where
\begin{equation}
n\bar{n} \equiv (u\bar{u} + d\bar{d})/\sqrt{2}.
\end{equation}
The mixing angle $\alpha$ is related to the usual nonet mixing
angle $\theta$ \cite{PDG} by the relation
\begin{equation}
\alpha = 54.7^{\circ} + \theta.
\end{equation}
For $\theta=0$ the quarkonium state becomes pure SU(3)$_f$
octet, while for $\theta=\pm 90^{\circ}$ it becomes pure singlet.
Ideal mixing occurs for $\theta=35.3^{\circ}$ (-54.7$^{\circ}$)
for which the quarkonium state becomes pure $\ssbar$
($\nnbar$).

In general we define
\begin{equation}
\eta = {\rm cos}\phi |n\bar{n}\rangle - {\rm sin}\phi
|s\bar{s}\rangle
\end{equation}
and
\begin{equation}
\eta' = {\rm sin} \phi |n\bar{n}\rangle +
{\rm cos} \phi |s\bar{s}\rangle
\end{equation}
with $\phi = 54.7^{\circ} +\theta_{PS}$, where $\theta_{PS}$ is the
usual octet-singlet mixing angle in SU(3)$_f$ basis where
\begin{equation}
\eta = {\rm cos}(\theta_{PS}) |\eta_8\rangle -
{\rm sin}(\theta_{PS}) |\eta_1\rangle,
\end{equation}
\begin{equation}
\eta' = {\rm sin}(\theta_{PS}) |\eta_8\rangle +
{\rm cos}(\theta_{PS}) |\eta_1\rangle.
\end{equation}

The decay of quarkonium into a pair of mesons
$ \qqbar \rightarrow M(Q\bar{q_i}) M(q_i\bar{Q})$
involves the creation of $q_i\bar{q_i}$ from the vacuum.
If the ratio of the matrix elements for the creation
of $s\bar{s}$ versus $u\bar{u}$ or $d\bar{d}$ is denoted by
\footnote{we shall assume that $\langle 0|V|d\bar{d}\rangle
\equiv \langle 0|V|u\bar{u}\rangle$.}

\begin{equation}
\rho \equiv \frac{\langle 0|V|s\bar{s}\rangle}{ \langle
0|V|d\bar{d}\rangle},
\end{equation}
then the decay amplitudes of an isoscalar $0^{++}$ (or $2^{++}$)
are proportional to

\begin{eqnarray}
\langle\qqbar|V|\pi\pi\rangle & = & {\rm cos} \alpha
\nonumber\\
\langle\qqbar|V|\KKbar\rangle & = & {\rm cos}
\alpha (\rho - \sqrt{2} {\rm tan}
\alpha)/2 \nonumber\\
\langle\qqbar|V|\eta\eta\rangle & = & {\rm cos}
\alpha (1 - \rho \sqrt{2} {\rm tan}
\alpha )/2 \nonumber\\
\langle\qqbar|V|\eta\eta^\prime\rangle & = & {\rm cos}
\alpha (1 + \rho \sqrt{2} {\rm
tan} \alpha )/2. \nonumber\\
\label{eq:quarkonium0}
\end{eqnarray}

The corresponding decay amplitudes of the isovector are

\begin{eqnarray}
\langle\qqbar|V|\KKbar\rangle & = & \rho/2 \nonumber\\
\langle\qqbar|V|\pi\eta\rangle & = & 1/\sqrt{2} \nonumber\\
\langle\qqbar|V|\pi\eta^\prime\rangle & = & 1/\sqrt{2},
\nonumber\\
\label{eq:quarkonium1}
\end{eqnarray}
and those for $K^*$ decay

\begin{eqnarray}
\langle\qqbar|V|K\pi\rangle & = & \sqrt{3}/2 \nonumber\\
\langle\qqbar|V|K\eta\rangle & =
& (\sqrt{2}\rho - 1)/\sqrt{8} \nonumber\\
\langle\qqbar|V|K\eta^\prime\rangle & = & (\sqrt{2}\rho + 1)
/\sqrt{8}. \nonumber\\
\label{eq:quarkonium2}
\end{eqnarray}

For clarity of presentation we have presented
eqn. \ref{eq:quarkonium0},\ref{eq:quarkonium1} and
\ref{eq:quarkonium2} in the approximation where
$\eta \equiv (n\bar{n} - s\bar{s}) / \sqrt{2}$ and
$\eta' \equiv (n\bar{n} +  s\bar{s}) / \sqrt{2}$, i.e. for a
pseudoscalar mixing angle $\theta_{PS} \sim -10^{\circ}$ ($\phi =
45^{\circ}$). This is a useful mnemonic;  the full expressions for
arbitrary $\eta, \eta'$ mixing angles $\theta_{PS}$ are given in
appendix A and are used in detailed comparisons throughout this
paper. Exact SU(3)$_f$ flavour symmetry corresponds to $\rho =
1$; empirically $\rho \geq 0.8$ for well established nonets such as
$1^{--}$ and $2^{++}$ \cite{dok95,Godfrey}.

The partial width into a particular meson pair $M_iM_j$ may be
written as
\begin{equation}
\Gamma_{ij} = c_{ij}|M_{ij}|^2\times |F_{ij} (\vec{q})|^2 \times p.s.
(\vec{q})
\equiv \gamma^2_{ij} \times |F_{ij} (\vec{q})|^2 \times p.s.
(\vec{q})
\label{Gamma}
\end{equation}
where $p.s.(\vec{q})$ denotes the phase-space, $F_{ij} (\vec{q})$
are model-dependent form factors, $M_{ij}$ is the relevant
amplitude (eqn. \ref{eq:quarkonium0},\ref{eq:quarkonium1} or
\ref{eq:quarkonium2}) and $c_{ij}$ is a weighting factor arising
from the sum over the various charge combinations, namely 4 for
$K\bar{K}$, 3 for $\pi\pi$, 2 for $\eta\eta^\prime$ and 1 for
$\eta\eta$ for isoscalar decay (eqn. \ref{eq:quarkonium0}), 4 for
$K\bar{K}$, 2 for $\pi\eta$ and 2 for $\pi\eta^\prime$ for
isovector decay (eqn. \ref{eq:quarkonium1}) and 2 for $K^*$
decays (eqn. \ref{eq:quarkonium2}). The dependence of
$\gamma^2_{ij} = c_{ij}|M_{ij}|^2$ upon the mixing angle $\alpha$
is shown in fig. \ref{alpha}a  for the isoscalar decay in the case of
SU(3)$_f$ symmetry, $\rho = 1$.

We confront the above with data on the established $2^{++}$
nonet, determine a probable range of values for $\rho$, and then
compare with $f_0(1500)$ decays.

\section{Flavour Symmetry in Meson Decays}
\subsection{$\qqbar$ Decays: the $2^{++}$ nonet}
Exact SU(3)$_f$ flavour symmetry requires the parameter $\rho$
to be unity. To get a feeling for symmetry breaking in the
$\qqbar$ sector we have computed some of the expected
branching ratios for the tensor mesons $a_2(1320)$, $f_2(1270)$
and $K^*_2(1430)$ decaying to two pseudoscalars and have
compared them with data \cite{PDG}. The decay branching ratio
$B$ of a $\qqbar$ state is proportional to the partial width (eqn.
\ref{Gamma}). We use for the phase space factor $p.s.(\vec{q}) =
q$ and for the formfactor
\begin{equation}
|F_{ij} (\vec{q})|^2 = q^{2\ell} \times exp(- q^2/8\beta^2),
\label{gamma2}
\end{equation}
where $\ell$ = 2 is the angular momentum in the final state with
daughter momenta $q$ and $\beta \simeq$ 0.4 GeV/c
\cite{Godfrey}. The ratios of the various partial widths are rather
insensitive to choice among different successful descriptions of
meson spectroscopy and dynamics. The detailed sensitivity to
form factors is discussed in appendix B.

The pseudoscalar mixing angle $\theta_{PS}$ has the empirical
value -$(17.3 \pm 1.8)^{\circ}$ \cite{AmslerPS} (and hence $\phi
= 37.4^{\circ}$, somewhat removed from the ``ideal" $45^{\circ}$
used in section 3). We shall use the full expressions given in
appendix A and the above value of $\theta_{PS}$ in all
phenomenology. Analogously, for the tensor mixing angle,
$\theta$ = 26$^{\circ}$ \cite{PDG}. The predictions are fitted to the
experimental values \cite{PDG}

\begin{eqnarray}
B(f_2\rightarrow\eta\eta)/B(f_2\rightarrow\pi\pi) =
(5.3 \pm 1.2) \times10^{-3}, \nonumber\\
B(f_2\rightarrow\KKbar)/B(f_2\rightarrow\pi\pi) = (5.4 \pm 0.6)
\times 10^{-2}, \nonumber\\
B(a_2\rightarrow\eta\pi)/B(a_2\rightarrow\KKbar) = 2.95
\pm 0.54, \nonumber\\
B(a_2\rightarrow\eta'\pi)/B(a_2\rightarrow\KKbar) = 0.116 \pm
0.029. \nonumber\\
\end{eqnarray}

Figure \ref{chivsrho} shows the $\chi^2$ distribution as a function
of $\rho$ for various values of $\beta$. The distribution
does not change significantly for $\beta > $1 GeV/c.  A good fit
(with a $\chi^2$ confidence level of more than 5\%) is obtained
with $\beta$ =  0.5 GeV/c (or larger) for which $\rho$ = 0.96
$\pm$ 0.04. This result is consistent with $K_2^*(1430)$ decays
although the experimental errors are large: $B(K^*_2\rightarrow
K\pi)/B(K^*_2\rightarrow K\eta)$ = ($3.6 ^{+7.1}_{-2.3})
\times10^{3}$ \cite{PDG}. We therefore conclude that flavour
symmetry breaking effects cannot be large in this established
$\qqbar$ nonet. Similar conclusions follow for a wide range of
$\qqbar$ decays (at least in the $\beta \rightarrow \infty$ limit
where form factors are ignored \cite{dok95}). Thus it seems
reasonable to expect that for $\qqbar$ scalar decays also,  $\rho
\approx 1$ and $\beta\sim  0.5$ GeV/c.

We have refrained from using the corresponding ratios for
$f_2'(1525)$ decays since the $\pi\pi$ decay width is (i) poorly
known and (ii) very sensitive to the precise tensor nonet mixing
angle $\theta$. For the ratio
$B(f_2'\rightarrow\eta\eta)/B(f_2'\rightarrow\KKbar)$ we find
0.07 for $\rho = 1$ and $\beta$ = 0.5 GeV/c, in good
agreement with experiments (0.11 $\pm$ 0.04) \cite{PDG} but at
variance with the value (0.39 $\pm$ 0.05) advocated by the
Particle Data Group which relies on one experiment only.

\subsection{The Decay Properties of $f_0$(1500)}

The branching ratios for $f_0(1500)$ production and decay
are \cite{Amsler3pi0,Enhan,Amsleretaeta}:
\begin{eqnarray}
B(\pbarp\rightarrow f_0\pi^0, f_0\rightarrow\pi^0\pi^0) =
(8.1 \pm 2.8) \times 10^{-4}, \nonumber\\
B(\pbarp\rightarrow f_0\pi^0, f_0\rightarrow\eta\eta) =
(5.5 \pm 1.3) \times 10^{-4}, \nonumber\\
B(\pbarp\rightarrow f_0\pi^0, f_0\rightarrow\eta\eta') =
(1.6 \pm 0.4) \times 10^{-4}. \nonumber\\
\end{eqnarray}
where the errors do not reflect the statistical significance
of the signals, but rather uncertainties  in the various assumptions
made in the fitting procedures. The decay branching ratios
are given by eqn. \ref{Gamma} with $p.s.(\vec{q}) = q$ and the
form factors
\begin{equation}
|F_{ij} (\vec{q})|^2 = {\rm exp}(- q^2/8\beta^2),
\label{ff}
\end{equation}
since $\ell$ = 0 for $0^{++}$ decays to two pseudoscalars.
The $f_0(1500)$ observed in $\eta\eta'$ decay has a mass of
1545 $\pm$ 25 MeV and lies just above threshold \cite{Enhan}.
We use for $q$ the average decay momentum (194 MeV/c)
derived from the damped Breit-Wigner function used in the
analysis of ref. \cite{Enhan}. The uncertainty in the
mass ($\pm$ 25 MeV) is taken into account when computing
the error on $\gamma^2$. With $\beta$ = 0.5 GeV/c
(section 4.1) we find:
\begin{equation}
R_1 =
\frac{\gamma^2(f_0(1500)\rightarrow\eta\eta)}{\gamma^2(f_0(1
500)\rightarrow\pi\pi)} = 0.27\pm 0.11,
\label{R1}
\end{equation}
\begin{equation}
R_2 =
\frac{\gamma^2(f_0(1500)\rightarrow\eta\eta')}{\gamma^2(f_0(1
500)\rightarrow\pi\pi)} = 0.19 \pm 0.08,
\label{R2}
\end{equation}
where $\pi\pi$ includes $\pi^+\pi^-$.  These results are in good
agreement with the results of the coupled channel analysis
\cite{coupled}. A signal for scalar decay to $\KKbar$ has not
yet been observed in $\pbarp$ annihilation in the the 1500 MeV
region. A bubble chamber experiment \cite{Gray} reports
$B.R.(p\bar{p} \rightarrow X\pi; X \rightarrow K\bar{K}) < 3.4
\times 10^{-4}$ which interpreted directly as an intensity leads to a
(90\% C.L.) upper limit:
\begin{equation}
R_3 =
\frac{\gamma^2(f_0(1500)\rightarrow\KKbar)}{\gamma^2
(f_0(1500)\rightarrow\pi\pi)} \stackrel{<}{\sim} 0.1.
\label{R3}
\end{equation}
However, interference effects among amplitudes could lead to an
underestimate for this number. We shall consider the implications
of the above $R_3$ value but shall also give results allowing for larger
values.

We find that if in the decay of some state the
ratios of partial widths (per charge combination and after phase
space and form factor corrections) for $\eta\eta$/$\pi \pi$ and
$\eta\eta/\KKbar$ are simultaneously both greater than unity,
then this state cannot be a quarkonium decay unless $\ssbar$
production is enhanced ($\rho > 1$) (see fig.
\ref{greaterthanone}).
The $f_0(1500)$
data on $\eta \eta / K\bar{K}$ satisfy this, but the
$\eta\eta/\pi\pi$ is inconclusive; at $1\sigma$ the ratio
per charge configuration gives $0.81 \pm 0.33$. If $f_0(1590)$ and
$f_0(1500)$ are the same state, as discussed above (see footnote 1), then if
the $\pi\pi$ branching ratio is reduced towards the GAMS
limit\cite{Gentral,Aldeall} the value of $R_1$ would rise such that it may
be possible to confirm the $f_0(1500)$ as a glueball by this test alone.
We cannot overemphasise the importance of a mutually consistent analysis
of the data on these experiments, in particular for clarifying the magnitude of
the $\eta\eta$/$\pi \pi$ ratio. If the ratio rises, as for GAMS, it would
immediately point towards a glueball; if the ratio remains as in eqn.\ref{R1}
then the arguments are less direct
but there still appears not to be a consistent
$Q\bar{Q}$ solution to the flavour
dependence of the ratios of partial widths and the magnitudes of the total
widths for the $f_0(1500)-f_0(1370)$ system.

Figure \ref{alpha} shows the invariant couplings $\gamma^2$ as a
function of $\alpha$ for $\rho = 1$ (a) and $\rho = 0.75$ and 1.25
(b), for a pseudoscalar mixing angle $\theta$ = -17.3$^{\circ}$
\cite{AmslerPS}.
The effects of SU(3)$_f$ breaking, $\rho < 1$, in the region where
$n\bar{n}$ dominates the Fock state ( $0 \leq \alpha \leq
30^{\circ}$) are interesting (fig. \ref{alpha}b).  We see  that the
branching ratios for $\eta$ or $\eta^\prime$ are little affected
(essentially because they are produced via the $\nnbar$
component in the $\eta$ which is $\rho$ independent) whereas
$K\bar{K}$ depends on $\rho$ (due to $s\bar{s}$ creation
triggering $K\bar{K}$ production from an $n\bar{n}$ initial state).
Thus we can suppress $\KKbar$ by letting $\rho\rightarrow 0$
without affecting the $\eta\eta/\pi\pi$ ratio substantially. In this
case the measured values for $\eta\eta$ and $\eta\eta'$ and the
upper limit for $\KKbar$ suggest that $\alpha\sim 0$, hence
$f_0(1500)$ is a pure $\nnbar$ meson. However this is still
unsatisfactory as the required value of $\rho$ implies a dramatic
suppression of $s\bar{s}$ creation to a degree not seen elsewhere
in hadron decays.

Figure \ref{rhotanalpha} shows the dependence of
$\rho$\ tan$\alpha$ (eqn. \ref{eq:quarkonium0}) on $R_1$ and
$R_2$. From the experimental values (eqn. \ref{R1} and \ref{R2})
we find consistency for $\rho$\ tan$(\alpha) \sim$  -0.1.
Hence either $\rho$ is small or $\alpha$ is small. The former
leads to unacceptable violation of SU(3)$_f$ and
the latter predicts the ratio
$\gamma^2(\KKbar)/\gamma^2(\pi\pi)$ to be 1/3.

Form factors  of the type in eqn. \ref{ff} tend to kill transitions at
large $q$, for example $F_{\eta\eta} (q_\eta)/F_{\pi\pi}
(q_\pi) > 1$  (i.e. opposite to naive phase space which grows
with  $q$ ). Without a form factor, e.g. $|F_{ij} (\vec{q})|\equiv 1$,
we obtain $R_1 = 0.31 \pm .13$, $R_2 = 0.25 \pm 0.11$, which
excludes a common range of $\rho$\ tan$(\alpha)$
(see fig. \ref{rhotanalpha}), and $R_3 < 0.12$.
The form factors used in ref. \cite{kokoski87} have a node at $q
\sim 0.9$ GeV/c and hence lead to an even stronger suppression
of the observed $\pi\pi$ intensity which dramatically reduces
$R_3$, in contradiction with the expected 1/3 for an $\nnbar$
state (see also appendix B).

In fig. \ref{rhovsalpha} we plot
the allowed regions of $\rho$ vs. $\alpha$. The grey area shows
the common values of $\rho$ and $\alpha$ which satisfy the
Crystal Barrel data each at the 90 \% C.L. while the black area
shows the restricted range allowed by $\KKbar$ (eqn. 20).

If one wishes to force $f_0 (1500)$ into a $\qqbar$
nonet, then independent
of form factors and SU(3)$_f$ breaking one is forced to
$\alpha\rightarrow 0$, whereby $f_0 (1500)$ has strong
$n\bar{n}$ content. This remains true even were the $K\bar{K}$ branching ratio,
currently being remeasured at LEAR,  significantly greater than the
$R_3$ value of eqn.\ref{R3}: the magnitude of the $K\bar{K}$/$\pi\pi$ ratio is
controlled more by flavour symmetry breaking than by the magnitude of the
$n\bar{n}$ - $s\bar{s}$ mixing angle in the $\alpha \rightarrow 0$ region.
 This immediately implies that the orthogonal
isoscalar member will be dominantly $s\bar{s}$, with mass above
1500 MeV and prominent in $K\bar{K}$.  The $f'_0(1525)$ if
confirmed or the ``$\theta$" ($f_J(1720)$), if $0^{++}$ , could be
this state. However, the $f_0 (1370)$, seen in both $\pi\pi,
\eta\eta$ is then left in isolation.

In the next sections we confront the data on $G = f_0$ (1500) with
the extreme hypothesis that it is dominantly a glueball.  Its
production in the canonical glue enhanced environments of
$J/\psi\rightarrow\gamma (G\rightarrow 4\pi)$ \cite{bugg}, $
pp\rightarrow p(G)p$ \cite{Kirk}, and $p\bar{p}$ annihilation
\cite{Anis,Amsler3pi0,Augustin,Enhan,Hasan1,Hasan2}
is consistent with this hypothesis, and
recent lattice QCD studies \cite{ukqcd,weing}
predict that a scalar glueball exists in this region of mass.
It thus scores well on two of the three glueball figures of
merit \cite{closerev}.

We shall now consider the dynamics and phenomenology of glueball decays.

\section{Primitive Glueball Decays}
The decays of $c\bar{c}$, in particular $\chi_{0,2}$, provide a
direct window on $G$ dynamics in the $0^{++},2^{++}$ channels
insofar as the hadronic decays are triggered by $c\bar{c}
\rightarrow gg \rightarrow Q\bar{Q}Q\bar{Q}$ (fig.
\ref{3graphs}a). It is necessary to keep in mind that these are in a
different kinematic region to that appropriate to our main
analysis but, nonetheless, they offer some insights into the gluon
dynamics. Mixing between hard gluons
and $0^{++}$, $2^{++}$ $Q\bar{Q}$ states (fig. \ref{3graphs}c)
is improbable at these energies as the latter 1 - 1.5 GeV
states will be far off their mass-shell. Furthermore,the
 narrow widths of $\chi_{0,2}$ are consistent with
the hypothesis that the 3.5 GeV region is remote from the
prominent
$0^+,2^+$ glueballs, $G$. Thus we
expect that the dominant decay dynamics is triggered by hard
gluons directly fragmenting into two independent $Q\bar{Q}$
pairs (fig. \ref{3graphs}a) or showering into lower energy gluons
(fig. \ref{3graphs}b). We consider the former case now; mixing
with $Q\bar{Q}$ (fig. 6c) and $G \rightarrow GG$ (fig. 6b) will be discussed in
section 6.

\subsection*{$ G \rightarrow QQ\bar{Q}\bar{Q}$}
This was discussed in ref. \cite{closerev} and
the relative amplitudes for the process shown in fig.
\ref{3graphs}a
read
\begin{eqnarray}
\langle G|V|\pi\pi\rangle & = & 1 \nonumber\\
\langle G|V|\KKbar\rangle & = &  R \nonumber\\
\langle G|V|\eta\eta\rangle & = & (1+  R^2)/2 \nonumber\\
\langle G|V|\eta\eta^\prime\rangle & = & (1- R^2)/2,
\nonumber\\
\label{b}
\end{eqnarray}
with generalizations for arbitrary pseudoscalar mixing angles
given in  appendix  A and where $R \equiv \langle
g|V|s\bar{s}\rangle/\langle
g|V|d\bar{d}\rangle$. SU(3)$_f$ symmetry corresponds to
$R^2=1$. In this case the relative branching ratios (after weighting
by the number of charge combinations) for the decays
$\chi_{0,2} \rightarrow \pi\pi,\eta\eta,\eta\eta',K\bar{K}$
would be in the relative ratios 3 : 1 : 0 : 4. Data for $\chi_0$ are in
accord with this where the branching ratios are (in parts per mil)
\cite{PDG}:

\begin{eqnarray}
B(\pi^0\pi^0) & = & 3.1 \pm 0.6 \nonumber\\
\frac{1}{2}B(\pi^+\pi^-) & = & 3.7 \pm 1.1 \nonumber\\
\frac{1}{2}B(K^+K^-) & = & 3.5 \pm 1.2  \nonumber\\
B(\eta\eta) & = & 2.5 \pm 1.1. \nonumber\\
\label{chidata}
\end{eqnarray}

No signal has been reported for $\eta\eta'$.
Flavour symmetry is manifested in the decays of $\chi_2$ also:
\begin{eqnarray}
B(\pi^0\pi^0 )& = & 1.1 \pm 0.3 \nonumber\\
\frac{1}{2}B(\pi^+\pi^-) & = & 0.95 \pm 0.50 \nonumber\\
\frac{1}{2}B(K^+K^-) & = & 0.75 \pm 0.55  \nonumber\\
B(\eta\eta) & = & 0.8 \pm 0.5, \nonumber\\
\label{chidata2}
\end{eqnarray}
again in parts per mil. The channel $\eta\eta'$
has not been observed either.  These results are natural as they
involve hard gluons away from the kinematic region where $G$
bound states dominate the dynamics. If glueballs occur at lower
energies and mix with nearby $Q\bar{Q}$ states, this will in
general lead to a distortion of the branching ratios from the
``ideal" equal weighting values above (a detailed discussion of this
follows in section 6.1), and also in causing significant mixing
between $n\bar{n}$ and $s\bar{s}$ in the quarkonium
eigenstates. Conversely, ``ideal" nonets, where the
quarkonium eigenstates are $n\bar{n}$ and $s\bar{s}$, are
expected to signal those $J^{PC}$ channels where the masses of the
prominent glueballs are remote from those of the quarkonia.

An example of this is the $2^{++}$ sector where the quarkonium
members are ``ideal". Data on glue in the $2^{++}$ channel,
and potential mixing of glue with  $n\bar{n}/
s\bar{s}$, may be probed by $J/\psi \rightarrow \gamma
+ f_2(1270)/f_2'(1525)$ which measures
the $gg \rightarrow n\bar{n}/s\bar{s}$ amplitude (fig.
\ref{3graphs}a) insofar as $J/\psi \rightarrow \gamma +gg$
mediates these channels. The branching ratios in parts per mil are

\begin{eqnarray}
\frac{1}{2}B(J/\psi \rightarrow \gamma f_2(1270)) & = & 0.69
\pm 0.07
\nonumber\\
B(J/\psi \rightarrow \gamma f_2(1525))& = & 0.63 \pm 0.1.
\nonumber\\
\label{psi}
\end{eqnarray}

Here again there is no sign of significant symmetry breaking.
Furthermore we note the ideal $n\bar{n}$ and
$s\bar{s}$ nature of the $2^{++}$, manifested both
by the masses and the flavour dependence of the branching ratios,
which suggests that $G$ mixing is nugatory in this channel.
These data collectively suggest
that prominent $2^{++}$ glueballs are not in the $1.2 -
1.6$ GeV region which in turn is consistent with lattice calculations where
the mass of the $2^{++}$ primitive glueball is predicted to be larger than 2
GeV. The sighting of a $2^{++}$ state in the glueball favoured central
production, decaying into $\eta\eta$ with no significant $\pi\pi$\cite{2170}
could be the first evidence for this state.
In view of our earlier remarks on the $\eta\eta$/$\pi\pi$
and $\eta\eta$/$K\bar{K}$ ratios being a potentially direct signature for
a glueball, we recommend that a detailed search now be made for this state in
$\eta\eta$ {\bf and} $K\bar{K}$ (and $\pi\pi$) channels in central production.

The phenomenology of the $J^{PC} = 0^{++}$ sector in the 1.2 - 1.6
GeV region is rather different to this: the $f_0(1500)$ -
$f_0(1370)$ system cannot be described within a $Q\bar{Q}$
nonet, nor do the decay branching ratios of the $f_0(1500)$
respect the flavour blindness of glue, (eqn. 19 and 20).
We shall now begin to focus on this problem.

It was shown earlier \cite{closerev} that violation of flavour
symmetry ($R^2 \neq 1$) leads to smaller $\KKbar/\eta\eta$ and
a finite $\eta\eta'$, at least if graph \ref{3graphs}a dominates
glueball decay. This follows immediately from eqn. \ref{b} or from
the generalized formulae given in appendix A. The contributions
of graph \ref{3graphs}a are shown in fig. \ref{3con}a as a function
of $R^2$. Graph \ref{3graphs}a becomes compatible with the
Crystal Barrel data and the small $\KKbar$ ratio if $|R| \sim 0.3$
\cite{amsler94} - a rather strong violation of symmetry which
might be suggestive of significant mixing between
$G$ and flavoured states. If $R_3$ were as large as $\frac{1}{3}$ (the
value for an $n\bar{n}$ state) an attempt to interpret as gluonium would
still require $|R|$ to be as small as $0.5$. We now show that mixing of $G$ and
$Q\bar{Q}$ is to be expected if the strong coupling picture of QCD,
as in the lattice, is a guide to their dynamics.

\section{$Q\bar{Q}$ and Glueball Decays in Strong Coupling QCD}
In the strong coupling ($g\rightarrow\infty$) lattice formulation
of QCD, hadrons consist of quarks and flux links, or flux tubes, on
the lattice. ``Primitive" $Q\bar{Q}$ mesons consist of
a quark and antiquark connected by a tube of coloured
flux whereas primitive glueballs consist of a loop of flux
(fig. \ref{Rivsx}a,b) \cite{paton85}. For finite $g$ these eigenstates remain a
complete basis set for QCD but are perturbed by two types of
interaction \cite{kokoski87}:

\begin{enumerate}
\item
$V_1$ which creates a $Q$ and a $\bar{Q}$ at neighbouring lattice
sites, together with an elementary flux-tube connecting them, as
illustrated in fig. \ref{Rivsx}c,
\item
$V_2$ which creates or destroys a unit of flux around any
plaquette (where a plaquette is an elementary
square with links on its edges), illustrated in fig. \ref{Rivsx}d.
\end{enumerate}

The perturbation $V_1$ in leading order causes decays of
$Q\bar{Q}$ (fig. \ref{Rivsx}e) and also induces mixing
between  the ``primitive" glueball $(G_0)$ and  $Q\bar{Q}$
(fig. \ref{Rivsx}f). It is perturbation $V_2$ in leading order that
causes glueball decays and leads to a final state consisting of
$G_0G_0$ (fig. \ref{Rivsx}g); decays into $Q\bar{Q}$ pairs occur at
higher order, by application of the perturbation $V_1$ twice.  This
latter sequence effectively causes $G_0$ mixing with $Q\bar{Q}$
followed by its decay. Application of $V_1^2$  leads
to a $Q^2\bar{Q}^2$ intermediate state which then turns into
colour singlet mesons by quark rearrangement (fig.
\ref{3graphs}a); application of $V_2$ would lead to direct
coupling to glue in $\eta, \eta'$ or $V_2\times V_1^2$ to their
$\qqbar$ content (fig. \ref{3graphs}b).

The absolute magnitudes of these various contributions require
commitment to a detailed dynamics and are beyond the scope
of this first survey. We concentrate here on their {\bf relative}
contributions to the  various two body pseudoscalar meson final
states available to $0^{++}$ meson decays.For $Q\bar{Q}
\rightarrow Q\bar{q}q\bar{Q}$ decays induced by $V_1$, the
relative branching ratios are given in eqn. \ref{eq:quarkonium0}
where one identifies

\begin{equation}
\rho \equiv \frac{\langle Q\bar{s} s\bar{Q} |V_1| Q\bar{Q}
\rangle }{\langle Q\bar{d} d\bar{Q}|V_1| Q\bar{Q}\rangle}.
\label{rho}
\end{equation}
The magnitude of $\rho$ and its dependence on $J^{PC}$ is a
challenge for the lattice. We turn now to consider the effect of
$V_1$ on the initial ``primitive" glueball $G_0$. Here too we allow
for possible flavour dependence and define

\begin{equation}
R^2 \equiv  \frac{\langle s\bar{s} |V_1| G_0 \rangle}
{\langle d\bar{d} |V_1| G_0 \rangle }.
\label{r2}
\end{equation}

The lattice may eventually guide us on this magnitude and also on
the ratio $R^2/\rho$. In the absence of this information we shall
leave $R$ as free parameter and set $\rho=1$.

\subsection{Glueball-$\qqbar$ mixing at $O(V_1)$}
In this first orientation we shall consider mixing between $G_0$
(the primitive glueball state) and the quarkonia, $n\bar{n}$ and
$s\bar{s}$, at leading order in $V_1$ but will ignore that between
the two different quarkonia which is assumed to be higher order
perturbation.

The mixed glueball state is then

\begin{equation}
G = |G_0\rangle + \frac{|n\bar{n}\rangle \langle
n\bar{n}|V_1|G_0 \rangle}{E_{G_0}-E_{n\bar{n}}}
+ \frac{|s\bar{s}\rangle \langle s\bar{s}|V_1|G_0 \rangle}{E_{G_0}-
E_{s\bar{s}}}
\label{perturb}
\end{equation}
which may be written as

\begin{equation}
G = |G_0\rangle + \frac{\langle
n\bar{n}|V_1|G_0\rangle}{\sqrt{2}(E_{G_0}-
E_{n\bar{n}})}\{\sqrt{2}
|n\bar{n}\rangle +\omega R^2 |s\bar{s}\rangle  \}
\end{equation}
where

\begin{equation}
\omega \equiv \frac{E_{G_0}-E_{n\bar{n}}}{ E_{G_0}-E_{s\bar{s}}}
\label{omega}
\end{equation}
 is the ratio
of the energy denominators for the $n\bar{n}$ and $s\bar{s}$
intermediate states in old fashioned perturbation theory (fig.
\ref{3graphs}d).

Denoting the dimensionless mixing parameter by

\begin{equation}
\xi \equiv \frac{\langle d\bar{d}|V_1|G_0 \rangle }{E_{G_0}-
E_{n\bar{n}}},
\end{equation}
the three eigenstates become, to leading order in the perturbation,

\begin{eqnarray}
N_G |G \rangle = |G_0 \rangle + \xi \{\sqrt{2}|n\bar{n}\rangle  +
 \omega R^2 |s\bar{s} \rangle \}
\equiv |G_0 \rangle + \sqrt{2}\xi |\qqbar\rangle \nonumber\\
N_s |\Psi_{s} \rangle \equiv |s\bar{s} \rangle - \xi R^2 \omega
|G_0 \rangle
\nonumber\\
N_n |\Psi_{n} \rangle \equiv |n\bar{n} \rangle - \xi \sqrt{2} |G_0
\rangle
\nonumber\\
\label{3states}
\end{eqnarray}
 with the normalizations
\begin{eqnarray}
N_G  = \sqrt{1 + \xi^2 (2 + \omega^2 R^4)} ,\nonumber\\
N_s  = \sqrt{1 + \xi^2 \omega^2 R^4)},
\nonumber\\
N_n = \sqrt{1 + 2 \xi^2}.
\nonumber\\
\end{eqnarray}

Recalling our definition of quarkonium mixing

\begin{equation}
|\qqbar \rangle  = {\rm cos} \alpha |n\bar{n} \rangle -
{\rm sin} \alpha |s\bar{s}
\rangle
\end{equation}
we see that $G_0$ has mixed with an effective quarkonium of
mixing angle $\alpha$ where $\sqrt{2}{\rm tan} \alpha = -
\omega R^2$ (eqn . \ref{eq:quarkonium0}). For example, if
$\omega R^2 \equiv 1$, the SU(3)$_f$ flavour symmetry
maps a glueball onto quarkonium where tan$\alpha = -
1/\sqrt{2}$ hence $\theta=-90^{\circ}$, leading to the familiar
flavour singlet
\begin{equation}
|\qqbar \rangle = |u\bar{u} +d\bar{d} +s\bar{s} \rangle /\sqrt{3}.
\end{equation}

When the glueball is far removed in mass from the $Q\bar{Q}$,
$\omega \rightarrow 1$ and flavour symmetry ensues;
the $\chi_{0,2}$ decay and the $2^{++}$ analysis of sections 4 and 5
are examples of this ``ideal" situation. However, when $\omega
\neq 1$, as will tend to be the case when $G_0$ is in the vicinity
of the primitive $Q\bar{Q}$ nonet (the $0^{++}$ case of interest
here), significant distortion from naive flavour singlet can arise.

If the $G_0$ component contributed negligibly to the decays, the
expectations would be that there is (eqn. \ref{3states}) (i)
a state $\Psi_n \rightarrow \pi\pi, \eta\eta,
K\bar{K}$ which is compatible with  $f_0(1370)$;  (ii) a state
$\Psi_s\rightarrow K\bar{K}, \eta\eta, \eta \eta'$, but not
$\pi\pi$, to be established; (iii) the 1500 MeV state $G$ for
which the decay amplitudes relative to $\pi\pi$ are (replacing
$\sqrt{2}$ tan$ \alpha$ by $-\omega R^2$ in eqn.
\ref{eq:quarkonium0}),

\begin{eqnarray}
\langle G|V|\pi\pi\rangle& = & 1 \nonumber\\
\langle G|V|\KKbar\rangle & = & (\rho + \omega R^2)/2
\nonumber\\
\langle G|V|\eta\eta\rangle & = & (1 + \omega \rho R^2)/2
\nonumber\\
\langle G|V|\eta\eta^\prime\rangle & = & (1 - \omega \rho R^2)/2,
\nonumber\\
\label{a}
\end{eqnarray}
with generalisation given in appendix A. The invariant decay
couplings $\gamma_{ij}^2$ are shown in fig. \ref{3con}c as a function
of $\omega R^2$ for $\rho=1$. Thus, for example,
SU(3)$_f$ may be exact for the glue-quark coupling $(R=1)$ but
mass breaking effects ($\Delta m \equiv m_s-m_d
\neq 0$) can cause dramatic effects if $E_G-E_{n\bar{n}}$ or
$E_G-E_{s\bar{s}}$ is accidentally small, such that $\omega
\rightarrow 0$ or $\infty$ respectively.  We now consider an
explicit mixing scheme motivated by three mutually consistent
phenomenological inputs:
\begin{enumerate}
\item
The suppression of $K\bar{K}$ in the $f_0(1500)$ decays suggests
a destructive interference between $n\bar{n}$ and $s\bar{s}$
such that $\omega R^2 < 0$ (see fig. \ref{3con}c). This
arises naturally if the primitive glueball mass is
between those of $n\bar{n}$ and the primitive $s\bar{s}$. As the
mass of $G_0 \rightarrow m_{n\bar{n}}$ or $m_{s\bar{s}}$, the
$K\bar{K}$ remains suppressed though non-zero; thus eventual
quantification of the $K\bar{K}$ signal will be important.
\item
Lattice QCD suggests that the ``primitive" scalar glueball $G_0$
lies at or above 1500 MeV, hence above the $I=1$ $Q\bar{Q}$
state $a_0(1450)$ and the (presumed) associated $n\bar{n}$
$f_0(1370)$. Hence $E_{G_0}-E_{n\bar{n}} >0$ in the numerator of
$\omega$.
\item
The $\Delta m = m_{s\bar{s}} - m_{n\bar{n}} \approx 200-300$
MeV suggests that the primitive $s\bar{s}$ state is in the region
1600-1700 MeV. This is consistent with the requirement from (1)
and (2) that $m_{n\bar{n}} <m_{G_0} < m_{s\bar{s}}$.
\end{enumerate}

Higher order perturbation effects will be required
for a complete treatment, in particular including mixing between
$n\bar{n}$ and $s\bar{s}$, but that goes beyond this first
orientation and will require more data to constrain the analysis.
We shall present a posteriori evidence supporting
this leading order approximation.

Tests of this scenario and its further development will follow as
the predicted states are isolated and the flavour dependence of
their branching ratios is measured. In order to compute the decay
branching ratios of the physical (mixed) states, we need to
incorporate the contributions from the primitive glueball
components, $G_0$. We consider this now.

\subsection{$G_0 \rightarrow G_0G_0$ at $O(V_2)$}
Here the glueball decays directly into pairs of glueballs or mesons
whose Fock states have strong overlap with $gg$ (fig.
\ref{3graphs}b). This topology will not feed final states such
as $\pi\pi$ nor $K\bar{K}$ since gluons are isoscalar. To the
extent that there is significant $G$ coupling to $\eta,\eta^\prime$
or to  the $\pi\pi$ $S$-wave, $(\pi\pi)_s$, (e.g. $\psi' \rightarrow
\psi \eta$ and $\psi (\pi\pi)_s$ each have large intrinsic
couplings notwithstanding the fact that they are superficially OZI
violating) one may anticipate $\eta\eta,\eta\eta^\prime$, and
$(\pi\pi)_s (\pi\pi)_s$ in
the decays of scalar glueballs. Analogously for $0^{-+}$ glueballs
one may anticipate $ \eta(\pi\pi)_s$ or $\eta'(\pi\pi)_s$ decays.

The manifestation of this mechanism in final states involving
the $\eta$ or $\eta'$ mesons depends on the unknown overlaps
such as $\langle gg|V|q\bar{q} \rangle$ in
the pseudoscalars. We consider various possibilities from the
literature without prejudice at this stage.

In the limit $m_{u,d}\rightarrow 0$
chiral symmetry suggests that the direct coupling of glue to the
$\eta$ or $\eta'$ occurs dominantly through their $s\bar{s}$
content, thereby favouring the $\eta'$. This argument has been
applied to the $\eta(1460)$ in refs.\cite{ish,chan,goun88}:

\begin{equation}
\frac{\langle gg|V|\eta' \rangle}{\langle gg|V|\eta \rangle}
 = \frac{\langle \ssbar|\eta' \rangle}{\langle\ssbar|\eta \rangle} =
\frac{cot(\phi) + \sqrt{2} \lambda }
{\lambda \sqrt{2}cot(\phi) - 1},
\label{chir}
\end{equation}
where
\begin{equation}
\lambda \equiv \frac{\langle gg(0^-)|V|d\bar{d}\rangle}
{\langle gg(0^-)|V|s\bar{s} \rangle} \rightarrow 0
\end{equation}
in the chiral limit, for which the ratio in eqn. \ref{chir}
is $\sim -4/3$.

The ratio eqn. \ref{chir} depends sensitively on the pseudoscalar
mixing angle and on a small breaking of chiral symmetry but
remains negative in the range $-0.9 < \lambda < 0.5$. Thus we
anticipate
\begin{equation}
r_0 \equiv \frac{\langle \eta \eta'|V|G_0 \rangle}{\langle \eta\eta
|V|G_0\rangle} =
\frac{\langle gg|V|\eta' \rangle}{\langle gg|V|\eta \rangle}
\sim - \frac{4}{3}.
\label{chir1}
\end{equation}

There is some ambiguity as to how this is to be applied
quantitatively since $m_{\eta'} \neq m_{\eta}$ and the
wavefunctions at the origin $\psi_{\eta'}(0)$ and $\psi_{\eta}(0)$
are in general different. An alternative measure \cite{gr}
may be the ratio $\Gamma(\psi \rightarrow
\gamma \eta')/\Gamma(\psi \rightarrow
\gamma \eta) = 5.0 \pm 0.6$. Dividing out phase space factors $\sim p_\gamma^3$
we obtain the ratio of matrix elements
\begin{equation}
\label{jpsi}
r_0 (J/\psi) = \pm (2.48 \pm 0.15).
\end{equation}
The solution with the negative sign is compatible with a small
breaking of chiral symmetry ($\lambda = 0.18$).
This gives
similar results to arguments based on the gluon anomaly in the
pseudoscalar channel (see eqn. (60) in ref.\cite{dyak} and also
ref.\cite{frere}).

This enhanced gluonic production of $\eta\eta, \eta\eta'$
does not appear to be dramatic in the
$\chi_{0,2}$ decays as the $\eta\eta/\pi\pi$ ratio appears to be
``canonical" in the sense of section 5. This may be because $\langle
gg|V|\eta\rangle$ and $\langle
gg|V|\eta '\rangle$ are hidden in the large errors on the present data
(in which case isolation of $\chi \rightarrow \eta ' \eta '$ at a Tau Charm
 Factory would be especially interesting).
Alternatively it may be $\langle gg|V|(\pi\pi)_s \rangle$ that is
important and hence $\chi\rightarrow 4\pi$ is the signal. Indeed
this channel is the biggest hadronic branching ratio for both
$\chi_0$ and $\chi_2$. It would be interesting to compare the
$\pi\pi$ spectra in these final states with those in
$\eta_c \rightarrow \eta' \pi \pi$ and $\eta \pi \pi$ which are
dominant modes in the $\eta_c$ decays and may also be signals
for this dynamics. High statistics from a Tau-Charm Factory may
eventually answer this question. The relative coupling strengths of
$\eta \pi \pi$ and $\eta' \pi \pi$ in decays of the glueball candidate
$\eta(1420)$ are also relevant here.

Note that for the decays of $J^{PC} = 0^{++}, 2^{++}$ \ldots states,
one will have a sharp test for a glueball
  if non-perturbative effects favour the direct
$\eta\eta, \eta\eta'$  decay path over mixing with
$Q\bar{Q}$ systems. A
state decaying to $\eta\eta,\eta\eta^\prime$ and/or $\eta'\eta'$
but not $\pi\pi$ nor $K\bar{K}$ cannot be simply $\qqbar$ since
$\rho$ and $\alpha$ cannot sensibly suppress $\pi\pi$ and
$\KKbar$ simultaneously - (see e.g. fig. \ref{alpha} and
\ref{greaterthanone}). GAMS has claimed states at 1590  MeV
 ($\eta\eta,\eta\eta'$) \cite{Gentral,Aldeall}, 1740 MeV
$(\eta\eta)$ \cite{GAMS1770} and 1910 MeV
$(\eta\eta^\prime)$ \cite{GAMS1910} with no strong signal seen
in $\pi\pi$ nor $K\bar{K}$. If the existence of any of these
enigmatic states $X$ is seen in other experiments such as
central production $pp \rightarrow p(X)p$ or $\pbarp$
annihilation, where $X \rightarrow
\eta\eta, \eta\eta'$ with no $\KKbar$ signal, this will be strong
evidence for the presence of $G_0$ in their wavefunctions.
The possible hints of $f_J(2100)$ in $\eta$ channels
\cite{Hasan2} and of $f_2(2170)$ in $\eta\eta$ \cite{2170},
  if confirmed, will put a high premium on searching
for or limiting  the $\KKbar$ branching ratios for these states.  A
particular realisation of these generalities is the model of ref.
\cite{gr}. Similar remarks apply to the decay of pseudoscalars
$\rightarrow \eta (\pi \pi)_s$ in contrast to
$\overline{K}(K\pi)_s$. This may be the case for the
$\eta(1440)$  and for the $\eta_c$, as discussed above.

\section{Application to Scalar Mesons around 1.5 GeV}
\subsection{Decays of $f_0(1500)$}
We shall now combine these ideas with the other result of section
6,
namely that at $O(V_1)$ the $G_0$ mixes with
$Q\bar{Q}$ with amplitude $\xi$ and that the resulting $Q\bar{Q}$
components decay as in section 3.

For simplicity of analysis we shall set $\rho=1$.
{}From eqn. \ref{3states} and \ref{a} we obtain for $G \equiv
f_0(1500)$:
\begin{equation}
\langle \KKbar|V_1|G\rangle = \frac{1+\omega R^2}{2}
\langle \pi\pi |V_1|G\rangle.
\end{equation}
Eventual quantification of $R_3$ may be translated into a value of
$\omega R^2$ (see fig. \ref{newfig1}).
We shall scale all decay amplitudes relative to that for
$\langle \pi\pi |V|G\rangle$ and see
what this implies for the $G_0$ decay amplitudes.  Thus

\begin{equation}
r_1\equiv\frac{\langle\eta\eta |V|G\rangle}{\langle\pi\pi
|V|G\rangle} =
\frac{\langle\eta\eta |V|G_0 \rangle}{N_G\langle\pi\pi
|V|G\rangle}
 + (\frac{1+\omega R^2}{2}) +{\rm cos} 2\phi (\frac{1-\omega
R^2}{2})
= \pm(0.90 \pm 0.20)
\label{a1}
\end{equation}
from $R_1$ (eqn. \ref{R1}) and
\begin{equation}
r_2\equiv\frac{\langle\eta\eta^\prime  |V|G\rangle}{\langle\pi\pi
|V|G\rangle}=
\frac{\langle\eta\eta^\prime  |V|G_0\rangle}{N_G\langle\pi\pi
|V|G\rangle}
 + {\rm sin} 2\phi (\frac{1-\omega
R^2}{2}) = \pm (0.53 \pm 0.11)
\label{a2}
\end{equation}
from $R_2$ (eqn. \ref{R2}),
from which we predict the ratio
\begin{equation}
r_0 = \frac{\langle \eta \eta'|V|G_0 \rangle}{\langle \eta\eta |
V|G_0\rangle} = \frac{2r_2-{\rm sin}2\phi (1-\omega R^2)}{2r_1-
{\rm cos}\phi (1-\omega R^2) -1 - \omega R^2}.
\end{equation}
The ratio $r_0$ is plotted in fig. \ref{allchiral} as a function of
$\omega R^2$. Note that this ratio is rather sensitive to the precise
value of the mixing angle $\phi$. The solution with the $+$ sign in
eqn. \ref{a1} and the $-$ sign in eqn. \ref{a2} (which we refer to
as the ``+ -" solution) agrees very well with $\omega R^2$ in the
range predicted by the current $\KKbar$ suppression (fig. \ref{newfig1})
and with radiative $J/\psi$ decay (eqn. \ref{jpsi}). Figure
\ref{allchiral} also suggests another possible
solution (``- -") compatible with the $\psi \rightarrow \gamma \eta
/ \eta'$ ratio.

In the particular limit
$\omega R^2 = -1$, the $\eta\eta$ decay mode would be driven dominantly
by $G_0$ decay. The smaller rate for $\eta\eta'$ decay observed
by Crystal Barrel would then be  due to \underline{destructive} interference
between $G_0$ decay and the admixture of quarkonium in the
wave function common to the ``+ -" and ``- -"
solutions. The amplitudes for $G_0$ decay (fig. \ref{3graphs}b)
are given in appendix A and the invariant couplings shown in
fig. \ref{3con}b as a function of $\lambda$.

\subsection{$f_0(1370)$}
The decay amplitudes for $\Psi_n \rightarrow \pi\pi$ and
$K\bar{K}$
will be those of an $n\bar{n}$ state such that,
\begin{equation}
\frac{\gamma^2(\Psi_n\rightarrow K\bar{K})}
{\gamma^2(\Psi_n \rightarrow \pi\pi)} \simeq
\frac{1}{3}
\end{equation}
Assuming that the lower state $\Psi_n$ is $f_0(1370)$
the decay to $\eta\eta^\prime$ is kinematically forbidden and so
the $\eta\eta$ decay  will be the only one immediately sensitive
to the predicted $G_0$ component in the $f_0(1370)$ Fock state.
We find from the decay branching ratios measured by Crystal
Barrel \cite{Amsler3pi0,Amsleretaeta} \footnote{The $<$ sign reflects
the fact that the measured branching ratio for $f_0(1370)$ decay
to $\pi\pi$ also includes some contribution from  $f_0(980)$.}

\begin{eqnarray}
B(\pbarp\rightarrow f_0(1370)\pi^0, f_0\rightarrow\pi^0\pi^0) <
(2.6 \pm 0.4) \times 10^{-3}, \nonumber\\
B(\pbarp\rightarrow f_0(1370)\pi^0, f_0\rightarrow\eta\eta) =
(3.5 \pm 0.7) \times 10^{-4}, \nonumber\\
\label{f01370}
\end{eqnarray}
after phase space and form factor corrections:
\begin{equation}
\frac{\gamma^2 (\Psi_n \rightarrow \eta\eta)}
{\gamma^2 (\Psi_n \rightarrow \pi\pi)} > 0.07
\label{epi}
\end{equation}
or
\begin{equation}
\frac{\langle\Psi_n |V|\eta\eta\rangle}{\langle\Psi_n
|V|\pi\pi\rangle} >
0.46.
\label{epia}
\end{equation}
On the other hand, eqn. \ref{3states} predicts

\begin{equation}
\frac{\langle\Psi_n |V|\eta\eta\rangle}{\langle\Psi_n
|V|\pi\pi\rangle} =
{\rm} {\rm cos}^2 \phi - \sqrt{2} \xi \frac{\langle
G_0|V|\eta\eta\rangle}{\langle n\bar{n}|V|\pi\pi \rangle},
\end{equation}
which reduces to cos$^2\phi$ = 0.63 for a pure $n\bar{n}$ state.
Solving for $\langle
G_0|V|\eta\eta\rangle$ and introducing into eqn. \ref{a1} one
finds with eqn. \ref{epia}
\begin{equation}
\xi^2 = \frac{{\rm cos} ^2\phi - 0.46}{2r_1 - (1+\omega R^2)
- {\rm cos} ^2\phi (1-\omega R^2)}
\end{equation}
The result eqn.\ref{epia} then implies that $|\xi|$ must be small
and prefers the +- solution
rather than the - - solution. One finds the 90\% C.L.
upper limit
\begin{equation}
|\xi| < 0.47
\label{estixsi}
\end{equation}
which a posteriori justifies the first order perturbation used in the
derivation of eqns. \ref{perturb} and \ref{3states}.

Further data analysis is now needed to quantify the experimental
ratio eqn. \ref{epi} and compare it in detail with the above.
In any event, the branching ratio of this $f_0(1370)$
state is consistent with $n\bar{n}$ dominance and hence further
isolates the $f_0(1500)$ as an exceptional state.

\subsection{The $f_0(1370)$ - $f_0(1500)$ system}
It should by now be clear that it is the combination of the two
siblings, $f_0(1370)$ and $f_0(1500)$, rather than either one on
its own that reveals the need for degrees of freedom beyond
$q\bar{q}$. We now illustrate how the branching
ratios and widths of the pair manifest this quantitatively.

In the quark model of ref. \cite{Godfrey} the widths of $^3P_0$
are qualitatively ordered as $\Gamma(\nnbar) >
\Gamma(\ssbar) >
\Gamma(a_0) \geq \Gamma(K^*)$. Empirically
$\Gamma(a_0)$ = 270 $\pm$ 40 MeV, $\Gamma(K^*)$ = 287
$\pm$ 23 MeV which supports this pair to be members of
the nonet and leads one to expect for their partners that
$\Gamma(\nnbar)\sim$ 700  MeV and $\Gamma(\ssbar)
\sim$ 500 MeV. In the flux tube model of ref.\cite{kokoski87}
 (see also appendix B) after normalising to the known widths of the
$2^{++}$ nonet one finds typically $\Gamma(K^*_0) > 200$ MeV and
$\Gamma(a_0) > 300$MeV in accord with data, and predict $\Gamma(f_0^{n\bar{n}})
> 500$MeV. That the $f_0^{n\bar{n}}$ width will be very broad is a rather
general
conclusion of all standard quark models (see also ref.\cite{abs});
 unitarisation effects
do not alter this conclusion \cite{nt}.

The $f_0(1500)$ width of  116 $\pm$ 17 MeV is clearly out
of line with this, being even smaller than the $K^*$ and $a_0$
widths. The {\bf total} width of $f_0(1370)$ is not yet well
determined, 200-700 MeV being possible
\cite{Amsler3pi0,Amsleretaeta} depending
on the theoretical model used in the analysis.

These suppressions of widths are natural in the $G-\qqbar$
mixing scheme as the presence of the $G_0$ component
dilutes the effect of the leading $n\bar{n}$ component:

\begin{equation}
\Gamma(\Psi_n) = \Gamma(n\bar{n})/(1 + 2\xi^2) > \frac{2}{3}
\Gamma(n\bar{n}).
\end{equation}
The  $f(1500) $ by contrast is, in our hypothesis, a glueball in
leading order and with $n\bar{n}, s\bar{s}$
components at $0(\xi)$ in perturbation.  The decay amplitudes for
$f(1500)$  are all at  $0(\xi)$, as shown by our analysis above.

Quantitative measures arise if we concentrate on the decays into
pseudoscalar pairs. The measurements from Crystal Barrel
\cite{Amsler3pi0,Amsleretaeta}
give the ratios of branching ratios for $G = f_0(1500)$ as
\begin{equation}
\frac{\eta\eta}{\pi\pi} = 0.23, \ \frac{\eta\eta'}{\pi\pi} = 0.07
\end{equation}
which implies that $B(G \rightarrow \pi\pi) = 0.7 F_2^G$
(where $F_2^G$ is the fraction of of two-body decays). With
$\Gamma_G = 116$ MeV this implies that
$\Gamma (G\rightarrow \pi\pi) = 28.2 F_2$
MeV per charge mode and hence, after dividing out phase space
and form factors (738 MeV/c and 0.755 respectively) we have the
reduced dimensionless measure

\begin{equation}
\tilde{\Gamma}(G \rightarrow \pi\pi) \equiv \frac{1}{3}
\gamma^2(G
\rightarrow \pi\pi)= 0.05 F_2^G.
\end{equation}

We now perform the same manipulations for the $f_0(1370)$. The
$\pi\pi$ decay appears to be the dominant two-pseudoscalar
decay mode (eqn. \ref{f01370}); any error on neglecting
$\eta\eta$ in the analysis is likely to be masked anyway by the
uncertainty on the total width which is not known to better than a
factor of 3.5 (spanning 200 - 700 MeV). Dividing out the phase
space (671 MeV/c) and form factors (0.798) as before, we form
the reduced measure per charge combination

\begin{equation}
\tilde{\Gamma}(f(1370) \rightarrow \pi\pi) \equiv \frac{1}{3}
\gamma^2(G
\rightarrow \pi\pi) =  (0.13 - 0.44)F_2^f.
\end{equation}

Hence the ratio of measures for the two states is

\begin{equation}
\frac{\tilde{\Gamma}(G \rightarrow \pi\pi)}{\tilde{\Gamma}(f
 \rightarrow \pi\pi)} < 0.4 \frac{F_2^G}{F_2^f}.
\end{equation}

Assuming that $R=1$ but allowing $\omega$ and $\xi$ to be free,
we expect this ratio to be given by

\begin{equation}
\frac{\tilde{\Gamma}(G \rightarrow \pi\pi)}{\tilde{\Gamma}(f
 \rightarrow \pi\pi)} = \frac{2 \xi^2 (1 + 2 \xi^2)}{1 +
\xi^2(2+\omega^2)}
\end{equation}
If $F_2^f$ is not small, then independly of $F_2^G$:

\begin{equation}
|\xi| < 0.52,
\label{xsirange}
\end{equation}
which is consistent with our earlier, independent, estimate in
eqn. \ref{estixsi}. The sum of the partial widths of the two
states for the $\pi\pi$  channels is
\begin{equation}
\Gamma(\Psi_n ) + \Gamma(G) \simeq \Gamma (n\bar{n}).
\end{equation}

Ref. \cite{Curtis} reports a strong $\sim 300$ MeV $4\pi$ signal in
the 1400 MeV region. It is unlikely that this signal is $f_0(1370)$
since the inelasticity in the $\pi\pi$ S-wave would be very large
($\sim$ 80 \%). However, given the uncertain dynamics in the $\pi\pi$
sector one must allow for this possibility in which case
our result eqn. \ref{xsirange} would break down.
On the other hand, a $4\pi$ contribution to $f_0(1500)$ decay
would decrease the upper limit for $|\xi|$. A more detailed
analysis is now warranted to verify if this further qualitative
indication is supported and to quantify the resonant contributions
$F_2^G/F_2^f$ and their effect on the analysis above.

\subsection{$f_0'(``1600"$)}
The mass of the $\Psi_s$ state, the value of $\omega R^2$ and of $m(G_0)$
are all related in our scheme and currently we can say no more than that the
1520 MeV to 1850 MeV range is possible for the mass of the
$m(\Psi_s)$.
A small value of $|\xi|$ would then require that $\Psi_s$ decays
essentially like an $s\bar{s}$ state, with couplings
\begin{equation}
\gamma^2 (\pi\pi:K\bar{K}:\eta\eta:\eta\eta') = 0:4:\frac{1}{2}:2.
\end{equation}

The decay of $f_0(1500)$ to $K\bar{K}$ will set the scale.  The amount of $G_0$
mixing depends rather sensitively on the value of $\omega$.
 However, a  general result is
that there will be \underline{constructive} interference between
$G_0$ and $s\bar{s}$ for the $\eta \eta'$ channel in either the
``+-" or ``- -" solutions
\begin{equation}
\frac{\gamma^2 (\Psi_s \rightarrow \eta\eta^\prime)}{\gamma^2
(\Psi_s
\rightarrow K\bar{K})} >
\frac{\gamma^2( s\bar{s}\rightarrow
\eta\eta^\prime)}{\gamma^2 (
s\bar{s}
\rightarrow K\bar{K})}  = \frac{1}{2}.
\end{equation}

If we take the widths of $K^*(1430)$ and $a_0(1450)$ as a guide
to normalize the width of the nonet in the Godfrey-Isgur
model \cite{Godfrey}, we would anticipate $\Gamma(\ssbar)
\sim$ 500 MeV at a mass of 1600 MeV. The $\Psi_s$
width will be suppressed relative to that of a pure $s\bar{s}$ due
to the glueball component (reflected in the normalization,
eqn. \ref{3states}) but the actual branching ratios may be
sensitively dependent on dynamics. In the model of ref.
\cite{kokoski87} the $K\bar{K}$ and $\eta\eta$ widths are
suppressed if $m(\Psi_s) \rightarrow 1800MeV$, see appendix B.
The importance of the $\eta\eta'$ channel appears to be a solid
prediction.

If the $f_J(1710)$ is confirmed to have a $J=0$ componenet in $K\bar{K}$
but not in $\pi\pi$, this could be a viable candidate for a
$G_0$- $s\bar{s}$ mixture, completing the scalar meson system
built on the glueball and the quarkonium nonet.

\section{The Scalar $\qqbar$ Mesons}
Based on the analysis of the previous sections we shall assume
that $f_0(1500)$ is mainly glue  and shall examine whether
a reasonable scalar $\qqbar$ nonet can be constructed with the
other scalar mesons, neglecting first the small glue
admixture in the two mainly $\qqbar$ isoscalars.

As we have seen, there are too many isoscalar $0^{++}$ mesons
to fit in the $\qqbar$ ground state nonet. A possible classification
of the scalar mesons is shown in table \ref{scalartable}.
The $a_0(980)$ and
$f_0(980)$ are interpreted as $\KKbar$ molecules, a hypothesis
that may be tested at DA$\Phi$NE \cite{cik}. The
$a_0(1450)$, $f_0(1370)$, $f'_0(1600)$ and $K_0^*(1430)$ are
the members of the ground state $\qqbar$ scalar nonet. Note that
the masses of the strange and $I$ = 1 members are similar;
this is also the case for some other nonets, in particular
4$^{++}$ \protect \cite{PDG}.

The Crystal Barrel and Obelix Collaborations at LEAR also report
the observation of a $0^{++}$ state decaying to $\rho^+\rho^-$ and
$\rho^0\rho^0$  in $\pbarp$ annihilation at rest into
$\pi^+\pi^-3\pi^0$ \cite{Curtis} and $\pbarn$ annihilation at rest
in deuterium into $2\pi^+3\pi^-$ \cite{Adamo} (see
also \cite{Gaspero}). Given that mass ($\sim$ 1350 MeV) and
width ($\sim$ 380 MeV) are compatible with $f_0(1370)$, one
might assume that $f_0(1370)$
has been observed here in its $\rho\rho$ decay mode. The
large $\rho\rho$ branching ratio points, however,  to a
large inelasticity in the $\pi\pi$ S-wave around 1400 MeV.
Thus $f_0(1370)$ may split into two states, a
$\qqbar$ decaying to $\pi\pi, \KKbar$ and $\eta\eta$  and
another state decaying to $\rho\rho$, possibly a molecule
\cite{Tornqvist}.

The GAMS meson $f_0(1590)$ \cite{Aldeall} decays to $\eta\eta'$ and
$\eta\eta$ with a relative branching ratio of 2.7 $\pm$ 0.8
\cite{PDG} which, given the large error, is consistent with
eqn. \ref{R1} and \ref{R2} for $f_0(1500)$ (see footnote 1).
 We shall assume that
$f_0(1590)$ is either identical to $f_0(1500)$ or that it is the
predicted $\ssbar$ state. We are therefore left with one nonet,
two to three molecules and the supernumerary $f_0(1500)$.

\begin{table}
\caption[]{The scalar mesons, their observed decay modes and
their suggested assignment.}
\begin{center}
\begin{tabular}{|l|l|l|l|}
\hline
Isospin & State & Decays & Nature \\
\hline
1 & $a_0(980)$ & $\eta\pi, \KKbar$ & $\KKbar$ molecule\\
0 & $f_0(980)$  & $\pi\pi, \KKbar$ & $\KKbar$ molecule \\
\hline
1 & $a_0(1450)$ & $\eta\pi$ & $\qqbar$ $0^{++}$ nonet \\
0 & $f_0(1370)$ & $\pi\pi, \eta\eta, \rho\rho$ & $\qqbar$
$0^{++}$ nonet \\
& $[f^\prime_0(``1600")]$ & $[\KKbar, \eta\eta, \eta\eta']$ &
$\qqbar$
$0^{++}$ nonet \\
1/2 & $K_0^*(1430)$ & $K\pi$ & $\qqbar$ $0^{++}$ nonet \\
\hline
0 & $f_0(1500)$ & $\pi\pi, \eta\eta, \eta\eta' , 4\pi^0$ & glueball
\\
\hline
\end{tabular}
\end{center}
\label{scalartable}
\end{table}

If the $\ssbar$ member lies at 1600 MeV then from the linear
mass formula
\begin{equation}
{\rm tg}^2 \theta = \frac{4K^*_0-a_0-3f^\prime_0}{3f_0+a_0-
4K^*_0},
\end{equation}
where the symbols denote the particle masses, one obtains for the
$\qqbar$ nonet the singlet-octet mixing angle
($59 ^{+12}_{-7})^{\circ}$ (or $121^{\circ}$)).
This is not far from ideal mixing, $\theta =
35.3^{\circ}$ (or 125.3$^{\circ}$) .
The relative decay rates of the $0^{++}$ $\qqbar$ mesons to two
pseudoscalars are given in appendix A and hence a further
consistency check on this proposed nonet may be applied when
the partial decay widths eventually become known. Note that
the experimental widths for the known $\qqbar$ members of the
proposed nonet (table \ref{scalartable}) are in good
agreement with predictions, see appendix B.

In the quark model, deviation from ideal mixing
is due to mixing transitions between $\uubar(\ddbar)$ and
$\ssbar$ which leads to a non-diagonal mass matrix in the flavour
basis. The mixing energy (non-diagonal elements) is
\begin{equation}
3A = f_0 + f'_0 - 2K^*_0
\end{equation}
which vanishes for ideal mixing. For our $0^{++}$ nonet one finds
37 MeV. The quark model also predicts the
Schwinger sum rule
\begin{equation}
(f_0+f'_0)\times (4K^*_0-a_0)-3f_0f'_0 -\frac{1}{3}(K^*-a_0)^2=
\frac{8}{3}(a_0-K^*)^2 \times {\rm cos}^2 \delta
\end{equation}
where cos$^2\delta$ is the fractional $\qqbar$ in the $f_0(1370)$
or
$f^\prime_0(``1600")$
wave function \cite{Closebook}. The fraction of glue
in $f_0(1370)$ is according to eqn. \ref{3states}:
\begin{equation}
{\rm sin^2} \delta = \frac{2\xi^2}{1+2\xi^2} < 0.33.
\label{Sch}
\end{equation}
Figure \ref{Schwing} shows how the
mixing angle $\delta$ in eqn. \ref{Sch}
varies as the function of the $a_0$ mass for
various $f_0(1370)$ masses.
A consistent result is indeed obtained when $m(f_0(1370)) \sim$
1400 MeV and $m(f_0(1600))$ = 1600 MeV. The fraction of glue
in $f_0(1600)$ depends on $\omega R^2$ (eqn. \ref{3states}).

\section{Conclusions}
We have argued that the properties of the $f_0(1500) -
f_0(1370)$ system are incompatible with them belonging to a
quarkonium nonet. We suggest that $f_0(1500)$ is prominently a
glueball mixed with the $\nnbar$ and $\ssbar$ quarkonia, the
$f_0(1370)$ being dominantly $\nnbar$.

We have shown that a reasonable scalar nonet can be built with
the newly discovered $a_0(1450)$ setting the mass scale,
and its width,  together with that of
the $K^*(1430)$, setting the scale of the nonet
widths which are within 25 \% of those expected by earlier quark
model calculations. The width of $f_0(1500)$ is
dramatically suppressed relative to these and the width
of the $f_0(1370)$ may be also somewhat suppressed.
These results are in line with these two states being partners in
a glueball - $\qqbar$ mixing scheme. It is this fortunate property
that has enabled the $f_0(1500)$ to show up so prominently in
several experiments where glueball channels are favoured.

Further supporting these arguments we have shown that
there appears to be no dramatic intrinsic
violation of flavour symmetry in decays
involving gluons in a kinematic region where $0^{++}$ or
$2^{++}$ glueball bound states are expected
to be negligible. The flavour dependence of $f_0(1500)$ decays
suggested that a significant mixing between $G$ and
$Q\bar{Q}$ states is distorting the branching ratios
in the $0^{++}$ sector. We argued that the observed decay
branching ratios could be due to the scalar
glueball expected in this mass range, mixing with the two nearby
$\qqbar$ isoscalars, one lying below, the other above $f_0(1500)$.
The partial decay widths of the lower state, $f_0(1370)$,
are consistent with a mainly $(\uubar +\ddbar)$ state.
Our hypothesis also implies that the (mainly) $\ssbar$ state lies in
the 1600 - 1700 MeV region.

The quantitative predictions of our analysis depend on the
suppression of $f_0(1500)$ decay to $\KKbar$. Thus
detailed study of $p\bar{p}\rightarrow\pi K\bar{K}$ can be
seminal (i) in confirming the $\KKbar$ suppression, (ii) in
confirming the $K$ (1430) $\rightarrow
K\pi$ and $a_0(1450)\rightarrow\eta\pi$ and $\KKbar$,
(iii) in quantifying the signal for $f_0$ (1370) and $f_0$ (1500)
and (iv) in isolating the predicted $\ssbar$ member of the nonet.

Clarifying the relationship between the Crystal Barrel
$f_0(1500)$ and the $f_0(1590)$ of GAMS is
important. Particular emphasis should be placed on the strength of
the $\pi\pi$ branching ratio and the ratio of branching ratios
$\eta\eta$/$\pi\pi$ in light of its potentially
direct significance as a test for
glueballs. If the $f_0(1550 \pm 50)$ becomes accepted as a
scalar glueball, consistent with the predictions of the lattice, then
searches for the $0^{-+}$ and especially the $2^{++}$ at mass
2.22 $\pm$ 0.13 GeV \cite{teper} may become
seminal for establishing the lattice as a successful calculational laboratory.

\vskip 0.2in
\noindent {\bf Acknowledgements}

We thank T. Barnes, M. Benayoun, K.Bowler,
D. Bugg, Y. Dokshitzer, J.M. Fr\`ere, G. Gounaris, A. Grigorian,
R.Kenway,  E. Klempt, H.J Lipkin, J. Paton, Y. Prokoshkin, S. Spanier, M.
Teper,
D. Weingarten, D.Wyler and B. Zou for helpful discussions.

\pagebreak
\section*{Appendix A: \\
Amplitudes for arbitrary pseudoscalar mixing angles}

The generalisations of the amplitudes for an arbitrary
pseudoscalar mixing angle $\phi$  are as follows:
\subsection*{Quarkonium decay}

\begin{eqnarray}
\langle\qqbar|V|\pi\pi \rangle& = & {\rm cos} \alpha
\nonumber\\
\langle\qqbar|V|\KKbar\rangle & = &  {\rm cos} \alpha (\rho
- \sqrt{2} {\rm tan}
\alpha)/2 \nonumber\\
\langle\qqbar|V|\eta\eta\rangle & = &  {\rm cos} \alpha ( {\rm
cos}^2
\phi - \rho \sqrt{2} {\rm tan} \alpha
{\rm sin}^2 \phi )
 \nonumber\\
\langle\qqbar|V|\eta\eta^\prime\rangle & = &   {\rm cos}
\alpha {\rm cos} \phi {\rm sin} \phi
(1 + \rho \sqrt{2} {\rm tan} \alpha  ) \nonumber\\
\end{eqnarray}
for $I=0$

\begin{eqnarray}
\langle\qqbar|V|\KKbar\rangle & = & \rho /2 \nonumber\\
\langle\qqbar|V|\pi\eta\rangle & = & {\rm cos} \phi
\nonumber\\
\langle\qqbar|V|\eta\eta^\prime\rangle & = & {\rm sin} \phi
\nonumber\\
\end{eqnarray}
for $I=1$ and

\begin{eqnarray}
\langle\qqbar|V|K\pi\rangle & = & \sqrt{3}/2 \nonumber\\
\langle\qqbar|V|K\eta\rangle & = & (\rho {\rm sin} \phi - {\rm
cos}
\phi/\sqrt{2})/\sqrt{2} \nonumber\\
\langle\qqbar|V|K\eta^\prime\rangle & = & (\rho {\rm cos} \phi +
{\rm sin} \phi/\sqrt{2})/\sqrt{2}
\nonumber\\
\end{eqnarray}
for $I=1/2$.

\subsection*{$ G \rightarrow QQ\bar{Q}\bar{Q}$ (graph
\ref{3graphs}a)}
\begin{eqnarray}
\langle G|V|\pi\pi\rangle & = & 1 \nonumber\\
\langle G|V|\KKbar\rangle & = &  R \nonumber\\
\langle G|V|\eta\eta\rangle & = & {\rm cos}^2 \phi  +  R^2 {\rm
sin}^2
\phi \nonumber\\
\langle G|V|\eta\eta^\prime\rangle & = & {\rm cos} \phi {\rm sin}
\phi (1- R^2)
\nonumber\\
\end{eqnarray}

\subsection*{$G \rightarrow GG$ (graph \ref{3graphs}b)}
\begin{eqnarray}
\langle G|V|\pi\pi \rangle & = & 0 \nonumber\\
\langle G|V|\KKbar\rangle & = & 0 \nonumber\\
\langle G|V|\eta\eta\rangle & = & ({\rm cos} \phi \sqrt{2}
- \lambda {\rm sin} \phi)^2
\nonumber\\
\langle G|V|\eta\eta'\rangle & = & ({\rm cos} \phi \sqrt{2}
- \lambda {\rm sin} \phi)
({\rm sin} \phi \sqrt{2} + \lambda {\rm cos} \phi)
\nonumber\\
\end{eqnarray}
where
\begin{equation}
\lambda \equiv \frac{\langle G(0^-)|V|d\bar{d}
\rangle}{\langle G(0^-)|V|s\bar{s}\rangle}
\end{equation} in the
pseudoscalar channel.

\subsection*{$G \rightarrow \qqbar$ (graph \ref{3graphs}c)}
\begin{eqnarray}
\langle G|V|\pi\pi\rangle & = & 1 \nonumber\\
\langle G|V|\KKbar\rangle & = &  (\rho + \omega R^2)/2
\nonumber\\
\langle G|V|\eta\eta\rangle & = & ({\rm cos}^2 \phi  +
\omega\rho R^2 {\rm sin}^2 \phi )
\nonumber\\
\langle G|V|\eta\eta^\prime\rangle & = & {\rm cos}
\phi {\rm sin} \phi (1- \omega\rho R^2)
\nonumber\\
\end{eqnarray}

\pagebreak
\section*{Appendix B: \\
Form Factors}

In the main body of the text we used rather simple forms for
form factors, eqs (14). In order to test sensitivity to these
assumptions we consider here the consequences of more
structured form factors as arise when the dynamical effects of
flux-tube breaking are included.  The result is that momentum
dependent multiplicative factors enter additional to those already
in eqn (14); the general structure is discussed
in table 2 and appendix B of ref\cite{kokoski87}.
  In the approximation where the light
hadron wavefunctions have the same scale parameter (the
quantity $\beta$ in eqn (14)) the structure of
$S-$wave decay amplitudes (as for $^3P_0\rightarrow ^1S_0 +
^1S_0$) is $$ S = (1-\frac{2q^2}{9\beta^2}) \exp (-
\frac{q^2}{12\beta^2})$$

The factor $\frac{q^2}{\beta^2}$ in parenthesis, which
was not present before, arises from the coupling of the $^3P_0$ of
the initial $Q\bar{Q}$ meson with the $^3P_0$ of the $q\bar{q}$ pair created by
the breaking of the flux-tube and which seed the decay.  From detailed
fits to meson spectroscopy and decays it is known that $\beta^{-2}
\simeq$ 5 to 6 GeV$^{-2}$ and so the multiplicative factor $(1-
\frac{2q^2}{9\beta^2}$) does not significantly affect our analysis
of the $f_0$(1500).

In this appendix we have followed ref\cite{kokoski87} in
imposing $\beta(^3P_0)$ = 0.5 GeV and have used a more modern
value \cite{isgw} of $\beta$ = 0.4 GeV for the
$^1S_0$ and $^3P_2$ states.  Barnes et al \cite{abs} have made a
similar analysis with $\beta\simeq$ 0.4 GeV as a preferred
overall value and find similar results.  The strategy is to use the
known $^3P_2$ decays to set the overall scale following the prescriptions
in table 2 and appendix B of ref.\cite{kokoski87}.

\subsection*{(1) $f^n_0$(1370)}

Using $\Gamma(f_2\rightarrow\pi\pi$) = 157 MeV yields, in MeV
\bea
\Gamma(f^n_0\rightarrow\pi\pi)  & = 270\pm 25\nonumber\\
K\bar{K} & = 195\pm 20 \nonumber\\
\eta\eta & = 95\pm 10\nonumber
\eea
where the $\eta$ is assumed to be a 50:50 mixture of $s\bar{s}$
and $n\bar{n}$. The $(1-\frac{2q^2}{9\beta^2})$ factor has
suppressed the $\pi\pi$ more markedly than
the $K\bar{K}$ and $\eta\eta$, hence leading to a larger
$\eta\eta/\pi\pi$ and $K\bar{K}/\pi\pi$ ratio than in the main text.
Nonetheless one sees that the $\Gamma(f^n_0)>>\Gamma(f^n_2$)
still arises in line with the conclusion that a ``narrow" $f_0$ width
is out of line with a $^3P_0$ quarkonium state.
  This conclusion is reinforced by
the expectation based on spin counting arguments that
$\gamma^2 (\rho\rho) >\gamma^2(\pi\pi)$ and hence that there
should be a non-negligible $\Gamma(f^n_0\rightarrow\rho(\pi\pi)_p)$
in addition to the two-body channels.

\subsection*{(2) $K^*_0$(1430)}

Using $\Gamma(K^*_2\rightarrow K\pi)\simeq$ 50 MeV or
$\Gamma(f_2)$ as above yields consistent similar results, namely
in MeV
\bea
\Gamma(K^*_0\rightarrow K\pi) & = & 200\pm 20\nonumber\\
K\eta & \simeq & 0\nonumber\\
K\eta^\prime & = & 15 \pm 20\nonumber
\eea

The $K\eta^\prime$ has a large coupling for physical
$\eta\eta^\prime$ mixing angles but the width is very sensitive to
phase space.  We note that the Particle Data Group\cite{PDG} allow
(7$\pm$10)\% ``non-$K\pi$" for the $K^*_0$ decay and we have
assigned this to $K\eta^\prime$.

These results should be compared with the experimental value
$\Gamma(K^*_0)$ = 287$\pm$23 MeV.

\subsection*{(3) $a_0$(1450)}

Using $\Gamma(a_2\rightarrow  K\bar{K}$) = 5.2$\pm$0.9 MeV
\cite{PDG} we obtain for the corresponding channel
$\Gamma(a_0\rightarrow  K\bar{K})
\simeq$ 110 MeV.  Then with physical $\eta\eta^\prime$ mixing
angles we have $\Gamma(\eta\pi)\simeq$ 90 MeV,
$\Gamma(\eta^\prime\pi) \simeq$ 80 MeV.
There is considerable uncertainty in these widths, however, since
if we were to normalise by $\Gamma(f_2\rightarrow\pi\pi$)
instead of by $\Gamma(a_2\rightarrow K\bar{K}$)  we would find
$\Gamma(a_0\rightarrow KK
:\eta\pi: \eta^\prime\pi)\simeq$ 200:160:140 MeV.  Reflecting
these uncertainties we can merely summarise by
\bea
\Gamma(a_0\rightarrow 0^-0^-)_{theory} = 390\pm 110\;
MeV\nonumber\\
\Gamma(a_0)_{\exp} = 270\pm40\; MeV.\nonumber
\eea

The general conclusion that $\Gamma(f^n_0) > \Gamma(a_0) \geq
\Gamma(K^*_0)$ holds true.  Furthermore we note that these
results require that if the $a_0$(1450)
seen by Crystal Barrel is indeed $^3P_0$ $(Q\bar{Q})$, then
comparable partial widths
are expected for all of $K\bar{K}, \eta\pi$ and $\eta^\prime \pi$.
Quantifying these experimentally will be an important piece of the
total strategy in clarifying the nature of these scalar mesons.

\subsection*{(4) $f^s_0$(1.6-1.8)}

The  $(1-\frac{2q^2}{9\beta^2})$ can have a dramatic effect in the
upper  part of this range of masses.  Using the
$\Gamma(f^n_2\rightarrow \pi\pi)$ or
$\Gamma(f^s_2\rightarrow K\bar{K})$ as normalisation, the mass
dependence of the partial widths of a pure $^3P_0(s\bar{s}$) are
as follows, in MeV

\begin{center}
\begin{tabular}{l|l|l|l}
& 1600 & 1700 & 1800\\\hline
$\Gamma(f^s_0\rightarrow K\bar{K})$ & 270 & 155 & 85\\\hline
$\eta\eta$ & 45 & 25 & 20\\\hline
$\eta\eta^\prime$ & 195 & 170 & 190\\\hline
\end{tabular}
\end{center}

The effect of the momentum node is clearly seen in the $KK$ and
$\eta\eta$ channels, whereas the $\eta\eta^\prime$ maintains its
strength.  There is nothing in this pattern of widths that supports
a $^3P_0(Q\bar{Q}$)  interpretation of $f_0$(1500).  A detailed
study of quarkonium decays with similar conclusions is in
ref\cite{abs}.

\pagebreak
\section*{Figure Captions}

\begin{figure}[h]
\vspace{1mm}
\caption[]{$\gamma_{ij}^2$ as a function of $\alpha$ for $\rho=1$
(a)
and $\rho=0.75$ or $1.25$ (b) for quarkonium decay (up to a common
multiplicative factor). Dotted line:
$\pi\pi$; dash-dotted line: $\KKbar$; dashed line:
$\eta\eta'$; solid line: $\eta\eta$.}
\label{alpha}
\end{figure}

\begin{figure}[h]
\vspace{1mm}
\caption[]{$\chi^2$ for $2^{++}$ quarkonium decay as a function of
$\rho$ for various values of $\beta$. The 5\% C.L. limit is shown
by the horizontal line.}
\label{chivsrho}
\end{figure}

\begin{figure}[h]
\vspace{1mm}
\caption[]{$\eta\eta/\pi\pi$ vs $\eta\eta/\KKbar$ invariant
couplings per charge combination for various values of $\rho \leq
1$. The grey region where both ratios are larger than one is not
accessible to $\qqbar$ mesons unless $\ssbar$ production is
enhanced.}
\label{greaterthanone}
\end{figure}

\begin{figure}[h]
\vspace{1mm}
\caption[]{$\rho$tan($\alpha$) as a function of $\eta\eta/\pi\pi$
(full curve) and $\eta\eta'/\pi\pi$ (dotted curve). The
vertical arrows show the experimental ranges of $R_1$ and
$R_2$ for $f_0(1500)$ decay, left with form factor, right without
form factor.}
\label{rhotanalpha}
\end{figure}

\begin{figure}[h]
\vspace{1mm}
\caption[]{$\rho$ as a function of $\alpha$. The full curves show
the 5 and 10 \% C.L. upper (+) and lower(-) limits dependence for
the
experimental value $R_1$, the dashed curves for $R_2$. The
dotted
curves give the boundaries for the experimental value $R_3$. The
grey
region shows the range allowed by the experimental data on
$f_0(1500)$ decay to $\pi\pi$, $\eta\eta$ and $\eta\eta'$. The
black
region includes in addition the bubble chamber upper limit for
$\KKbar$.}
\label{rhovsalpha}
\end{figure}

\begin{figure}[h]
\vspace{1mm}
\caption[]{Contributions to gluonium decay: $QQ\bar{QQ}$
(a), $GG$ (b), $\qqbar$ (c) and interpretation as $Q\bar{Q}$ mixing
(d)
involving the energy denominator $E_G-E_{Q\bar{Q}}$}
\label{3graphs}
\end{figure}

\begin{figure}[h]
\vspace{1mm}
\caption[]{Predicted decay rates $\gamma_{ij}^2$ (up to a common
multiplicative constant) (a) as a function of
$R^2$ for  $QQ\overline{QQ}$ decay, (b) as a function of
$\lambda$ for $GG$ decay and (c) as a function of $\omega R^2$
for $\qqbar$ decays with $\rho=1$. Solid line: $\pi\pi$; dashed
line: $\KKbar$; dash-dotted line: $\eta\eta$; dotted line:
$\eta\eta'$.}
\label{3con}
\end{figure}

\begin{figure}[h]
\vspace{1mm}
\caption[]{Glueballs, quarkonia and perturbations:
(a) primitive $Q\bar{Q}$ and (b) primitive glueball $G_0$ in flux tube
simulation of lattice QCD; perturbation $V_1$ (c) and $V_2$ (d);
the effect of $V_1$ on $Q\bar{Q}$ is shown in (e), and on $G$ is shown in
(f); the effect of $V_2$ on $G$ is shown in (g).}
\label{Rivsx}
\end{figure}

\begin{figure}[h]
\vspace{1mm}
\caption[]{$R_3$ versus $\omega R^2$.}
\label{newfig1}
\end{figure}

\begin{figure}[h]
\vspace{1mm}
\caption[]{Predicted ratio $r_0$ as a function of $\omega R^2$ for
$f_0(1500)$ decays. The solid lines show the four possible
solutions from the Crystal Barrel results $r_1$ and $r_2$ with the
corresponding signs. The range allowed by the experimental
errors is shown by the dashed lines for the ``+ -" and ``- -" solution.
The solutions ``+ -" and ``- -" are compatible with $\omega
R^2$ in the range allowed by the $\KKbar$ suppression and
radiative $J/\psi$ decay (area between the parallel lines, indicated by the
grey flashes). A small breaking of chiral
symmetry favors the ``+ -" solution.}
\label{allchiral}
\end{figure}

\begin{figure}[h]
\vspace{1mm}
\caption[]{Fractional contribution of $\qqbar$ in the $f_0(1370)$
wave function from the Schwinger sum rule as a function of $a_0$
mass for various $f_0(1370)$ masses, assuming the (mainly)
$\ssbar$ state to lie at 1600 MeV. The horizontal arrow shows the
experimental uncertainty in the $a_0$ mass and the vertical
arrow shows the range allowed for $|\xi| < 0.5$.}
\label{Schwing}
\end{figure}
\end{document}